\documentclass[aip,jcp,preprint,floatfix,nofootinbib]{revtex4-1}
\usepackage{amsmath}
\usepackage{amssymb}
\usepackage[caption=false]{subfig}
\usepackage{color}
\usepackage[dvips]{graphicx}
\PassOptionsToPackage{caption=false}{subfig}
\usepackage{epsf}
\usepackage[dvips]{graphics,graphicx}
\usepackage{sidecap}
\sidecaptionvpos{figure}{c}
\usepackage{amsmath,amssymb,bm,color,mathrsfs,wasysym}
\usepackage{epsfig}
\usepackage{amsfonts}
\usepackage{cancel}
\usepackage{ulem}
\usepackage{float}

\newcounter{saveeqn}
\newcommand{\bmathlet}{\setcounter{saveeqn}{\value{equation}}%
\stepcounter{saveeqn}\setcounter{equation}{0}%
\renewcommand{\theequation}{\mbox{\arabic{saveeqn}\alph{equation}}}}
\newcommand{\emathlet}{\setcounter{equation}{\value{saveeqn}}%
\renewcommand{\theequation}{\arabic{equation}}}
\newcommand{\be}{\begin{equation}}
\newcommand{\ee}{\end{equation}}
\newcommand{\bea}{\begin{eqnarray}}
\newcommand{\eea}{\end{eqnarray}}
\newcommand{\bi}{\begin{itemize}}
\newcommand{\ei}{\end{itemize}}

\newcommand{\dsp}{\displaystyle}
\newcommand{\cR}{\color{red}\footnotesize\singlespacing}

\newcommand{\etal}{\textit{et al }}
\newcommand{\Sm}{{\rm Sm}}
\newcommand{\meanJ}{{\cal J}}
\usepackage[usenames,dvipsnames]{xcolor}
\def\singlespacing{\baselineskip=12pt}

\begin{document}

\title{Molecular field theory for biaxial smectic A liquid crystals}

\author{T.B.T. To}
\affiliation{School of Mathematics, University of Southampton, Southampton, SO17 1BJ, United Kingdom}

\author{T.J. Sluckin}
\email{T.J.Sluckin@soton.ac.uk}
\affiliation{School of Mathematics, University of Southampton, Southampton, SO17 1BJ, United Kingdom}

\author{G.R. Luckhurst}
\affiliation{School of Chemistry, University of Southampton, Southampton, SO17 1BJ, United Kingdom}

\date{\today}

\begin{abstract}
Thermotropic biaxial nematic phases seem to be rare, but biaxial smectic A phases less so.  Here we use molecular field theory to study a simple two-parameter model, with one parameter promoting a biaxial phase and the second promoting smecticity.  The theory combines the  biaxial Maier-Saupe  and McMillan models.  We use alternatively the Sonnet-Virga-Durand (SVD) and geometric mean approximations (GMA) to characterize molecular biaxiality by a single parameter.   For non-zero smecticity and biaxiality, the model  always predicts a ground state biaxial smectic A phase. For a low degree of smectic order, the phase diagram is very rich, predicting  uniaxial and biaxial nematic  and smectic phases, with in addition a variety of tricritical and tetracritical points. For higher degrees of smecticity, the region of stability of the biaxial nematic phase is restricted and eventually disappears, yielding to the biaxial smectic phase. Phase diagrams from the two alternative approximations for molecular biaxiality are similar, except inasmuch that SVD  allows for a first order isotropic-nematic biaxial transition, whereas GMA predicts a Landau point separating isotropic and biaxial nematic phases. We speculate that the rarity of  thermotropic biaxial nematic phases is partly a consequence of  the presence of stabler analogous smectic phases.
\end{abstract}


\maketitle

\section{Introduction}

The possible existence of  a biaxial nematic liquid crystal phase ($N_B$) was predicted over forty years ago by Freiser \cite{freiser70}.  In the intervening period the subject has been the focus of much intensive   theoretical (e.g. \cite{Boccara1976, RemlerHaymet1986, Rosso2007}), computational (e.g. \cite{berardi2008})  and experimental  (see e.g. \cite{Luckhurst2004,Madsen2004,Kumar2004}) research.  There are well-attested observations of lyotropic biaxial nematic phases \cite{YuSaupe1980}, but there has been considerable controversy over the well-publicized reports \cite{Madsen2004,Kumar2004} of the existence of thermotropic biaxial nematic phases (see e.g. \cite{GoertzGoodby2005, Senyuk2010, Kim2013}).  It is by now clear that at the very least  thermotropic biaxial nematic phases are rather rare. 

Why this is the case is not entirely clear. Most nematogenic molecules are to some extent biaxial, and those explicitly constructed with a view to observing the biaxial nematic phase rather more so.   One suggestion is that merely the temperature range over which the biaxial phase might be expected is rather narrow, while others have suggested that in lyotropics (but not in thermotropics), polydispersity might stabilize otherwise unstable phases \cite{Galindo2003}.  However, these suggestions are difficult to evaluate,  given the lack of suitable experimental points of reference.    

However, one possibility with some experimental support is that biaxial nematogens are also suitably shaped so that they may also form smectic phases.  The idea here is that the smectic biaxial phase $\Sm A_B$ would in some sense destabilize the $N_B$  phase (we also use the  notations $N_U, \Sm A_U, I$  to denote  the uniaxial nematic, unaixial smectic and isotropic liquid phases, respectively).  The first suggestion of a biaxial smectic phase was made by de Gennes (see Teixeira \textit{et al.} \cite{Teixeira2006}). The first detailed theory was developed by Matsushita \cite{Matsushita1981}, but his results have proved  difficult to interpret in a more general context. 

Experimentally, biaxial smectic phases are much less elusive than their nematic biaxial counterparts, giving some support to the intuitive idea of  a competition between smectic  and nematic biaxial stability.  For example, non-polar biaxial smectic A phases have been discovered in a binary mixture of a board-like mesogen with a board like non-mesogenic compound \cite{Hegmann2001, Hegmann2007},  and in compounds of dimers of a rod-like and a bent-core mesogenic units \cite{Yelamaggad2004,Wang2012}. These experiments claim to find two phase sequences:  $\mathrm{SmA}_B-N_B-I$ \cite{Yelamaggad2004} and $\mathrm{SmA}_B-N_U-I$ \cite{Wang2012}.  In addition, in a variety of different chemical contexts related biaxial smectic A phases with antiferroelectric order, usually denoted by $\Sm A_{db}$ or $\Sm A_dP_a$ ,  have also been observed    \cite{Eremin2001, ReddyScience2011,Sadashiva2002, Sadashiva2004,Yelamaggad2006}. In these materials, the phase sequence is always $\Sm A _{db}-\Sm A_U-I$. 

In two experimental cases  \cite{Yelamaggad2004,Wang2012}, the non-polar biaxial $\Sm A_B$ phase is formed directly from the nematic phase on lowering temperature. Interestingly, the phase sequence $\mathrm{SmA}_B-N_U-I$  \cite{Wang2012} has also been found theoretically  in the Matsushita molecular field theory \cite{Matsushita1981}, and also computationally,  in a set of Gay-Berne simulations for orthogonal paralellepiped molecules which favour face-to-face interaction \cite{ZannoniBSmA2000}. Here the $\mathrm{SmA}_B$ is formed directly from $N_U$.  By contrast, one might naively expect that  an intermediate $N_B$ phase would interpose itself between  the $\mathrm{SmA}_B$  and the $N_U$ phases.  The lack of intemediate $N_B$ gives further intuitive support to the idea that the $\mathrm{SmA}_B$ is preempting  the $N_B$,  and thus rendering it unobservable.  \\
\indent In this  paper we seek to develop simple molecular criteria which control the relative stability of the $N_B$ and $\Sm A_B$.  Our development builds on theories of the uniaxial smectic phase $\Sm A_U$  and of the  biaxial nematic phase $N_B$.  The basic theory of  the uniaxial smectic phase $\Sm A_U$ is that of  McMillan \cite{McMillan1971, McMillan1972}.  In  the current work, we also  decouple the spatial modulations from the orientational distribution in the full density-orientational distribution function, following  Kventsel \etal \cite{KLZ1985}. In order to provide  a tractable theory, in our treatment of biaxiality,  we reduce the number of parameters controlling the biaxiality from two to one. In order to include  a generic  set of phase diagrams, we employ two variants of this strategy; these have previously been  used in the context of nematic biaxiality.  The approximations are:  the geometric mean approximation (GMA)  \cite{Luckhurst1975},  and the Sonnet-Virga-Durand approximation (SVD)\cite{Virga2003}.  We  explain these key approximations in greater detail below.  Our theory also builds on previous work by Teixeira \etal  \cite{Teixeira2006}, who have sought to establish simple criteria for the formation of biaxial smectic phases formed from board-like molecules. \\
\indent
The paper is organized as follows.  In section \ref{sec:background}, we give a brief resume of previous theories of the smectic phase and of nematic biaxiality, emphasizing   the key approximations we shall employ later.  In section \ref{sec:MFTsection},  we then use these ideas to develop our present theoretical framework.    In section \ref{sec:results}, we present  results from our calculations. Finally in section 
\ref{sec:discussionsection} we discuss the context in which our results should be understood. 

\section{Theoretical background}
\label{sec:background}

\subsection{Basic material}
\label{subsec:basic}

In this paper, molecular orientation $\Omega$ is parameterized in terms of three Euler angles $\psi,\theta,\phi$ which take their conventional meanings (polar angle $\theta$ and zenithal angles $\psi, \phi$). The Euler angle $\psi$ is only of interest in the specifically biaxial phases.  We suppose constituent molecules to possess  $D_{2h}$ symmetry (i.e. that of an orthogonal parallellepiped). In some cases the molecules of  interest possess other symmetries, such as, for example,   the $C_{2v}$ symmetry occurring in bent-core molecules. However, in the cases of interest the second order tensor invariants, the means of which provide the order parameters in the theory,  are those given by the $D_{2h}$ symmetry.

The underlying basic theory, on all of our work relies, is the Maier-Saupe theory of nematic ordering (see e.g.  de Gennes and Prost \cite{deGennes}), for which the free energy per particle  is given by:
\be
\label{eq:MaierSaupe}
A=k_BT\int_\Omega d\Omega f(\Omega)\ln 4\pi f(\Omega) -\frac{1}{2}\epsilon S^2,
 \ee
where $S$ is the nematic order parameter (discussed in more detail in eq.\eqref{eq:op1} below), $k_B$ denotes the Boltzmann constant, $T$ is absolute temperature and $\epsilon$  sets the scale of the molecular anisotropic interaction.

In this paper, thermodynamic quantities, namely the free energy, internal energy, potentials of mean torque, distribution functions and partition functions are formulated per particle. In addition, the temperature and energy quantities are expressed in non-dimensional units. We denote $T$ as the temperature, $A$ as the free energy, $U$ as the internal energy and $U(z,\Omega)$ as the potential of mean torque. We shall work, where possible, in  non-dimensional quantities,  given in terms of their physical quantities by
\begin{align}
&T^*=\frac{k_BT}{\epsilon}; \quad 
A^* = \frac{A}{\epsilon}; \quad \nonumber \\
&U^* = \frac{U}{\epsilon}; \quad
U^*(z,\Omega) = \frac{U(z,\Omega)}{\epsilon},
\end{align}
where $U^*(z,\Omega)$ is a potential of mean torque on a molecule with orienation $\Omega$ at point $z$. 

Our theory of the $\Sm A_B$ phase combines elements of standard theories of the uniaxial smectic phase $\Sm A_U$ \cite{McMillan1971,McMillan1972, KLZ1985} and of the biaxial nematic phase $N_B$ \cite{Straley1974,Luckhurst1975,Boccara1976,RemlerHaymet1986,Virga2003}. In order to make our work self-contained, in this section we give  a brief account of these theories.
\ \\
\subsection{McMillan theory of smectics}
\label{subsec:smecticbackground}

The McMillan theory \cite{McMillan1971,McMillan1972} restricts itself to  uniaxial (i.e. $D_{\infty h}$) molecules, for which the only relevant Euler angle is the polar angle $\theta$.  The smectic layer thickness $d$ is given in this theory. There is no attempt at self-consistency in determining, for example, the thickness $d$ in terms of moments of a full interparticle potential, although other theories \cite{lipkin83,singh2000} do make this attempt.  An extra simplifying element (see eq.\eqref{eq:KLZseparation} below) was introduced by Kventsel \etal \cite{KLZ1985}.  There is a single non-dimensional parameter $\alpha$ which drives  smectic order. Increasing  $\alpha$ increases the smectic ordering temperature. A brief summary of the theory follows. 

The orientational-translational densities are given by a function 
\be
\label{eq:distfn1}
\rho(z,\theta)=\bar{\rho}d F(z,\theta),  
\ee
where $\bar{\rho}$ is the mean particle number density, and $F(z,\theta)$ is a normalized   orientational-translational distribution function. We suppress the variables $\psi$ and $\phi$. The former has no meaning for a $D_{\infty h}$ particle. We integrate over the latter, which is in any case irrelevant as the distribution function is independent of $\phi$.  $F(z,\theta)$ is periodic in $z$ with period $d$, and is independent of the transverse position within the layer. The normalization is such that 
\be
\label{eq:normalization1}
\int_0^{\pi} \sin{\theta} d\theta \int_0^d dz F(z,\theta) =1.
\ee
    
The distribution function $F(z,\theta)$ is factorized into pure translational and pure orientational distribution functions, using an approximation introduced by Kventsel \etal  \cite{KLZ1985}: 
\be
\label{eq:KLZseparation}
F(z,\theta)=f(\theta)g(z),
\ee
with $g(z)=g(z+d)$. This approximation in principle has only limited validity, but in practice has been used with a good degree of success (see e.g. \cite{Miyajima1990,Pardhasaradhi2013}). The functions $f(\theta)$ and $g(z)$  are each independently normalized:
\be
\label{eq:normalization2}
\int_0^{\pi}\sin\theta d\theta f(\theta)=\int_0^d g(z)dz =1
\ee

There are three order parameters. These are: 
\bmathlet
\begin{itemize}
\item[(a)] the orientational order parameter 
\begin{eqnarray}
\label{eq:op1}
S=\langle P_2(\cos\theta) \rangle =\left\langle \frac{1}{2}\left(3\cos^2\theta-1\right)\right\rangle \nonumber \\
= \int_0^{\pi} \sin{\theta} d\theta f(\theta) P_2(\cos\theta).
\end{eqnarray}
 the usual order parameter  entering the theory of nematics;
\item[(b)]  the smectic density wave amplitude:
\be
\label{eq:op2}
\tau = \left\langle \cos{ \left( \frac{2\pi z}{d} \right) } \right\rangle =  \int_0^d dz g(z) \cos\left(\frac{2\pi z}{d}\right).
\ee
\item[(c)]  the so-called mixed orientational-translational order parameter:
\begin{align}
\label{eq:op3}
&\sigma = \left\langle P_2(\cos\theta)\cos{ \left( \frac{2\pi z}{d} \right) } \right \rangle \nonumber \\
&= \int_0^{\pi} \sin{\theta} d \theta \int_0^d dz F(z,\theta) P_2(\cos\theta)\cos\left(\frac{2\pi z}{d}\right).
\end{align}
\end{itemize}
\emathlet

The distribution function decoupling eq.\eqref{eq:KLZseparation} reduces the number of independent order parameters from three to two, with
\be
\label{eq:op4}
\sigma=S\tau .
\ee

The Helmholtz free energy density relative to a disordered phase  is expressed in non-dimensional units as
\begin{align}
\label{eq:helmholtz1}
&A^*= T^* \int_0^{\pi}  \sin\theta d\theta f(\theta)\ln\Big(2f(\theta)\Big) \nonumber \\
&+ T^*\int_0^d dz g(z)\ln\Big(dg(z)\Big)-\frac{1}{2}\Big(S^2 +\alpha S^2\tau^2\Big).
\end{align}

The first term  comes from orientational entropy, the second from translational entropy, the third from orientational energy, while only the last term is due to the energy advantage of smectic order. The factors of 2 and $d$ in the entropy integrals are normalizing factors, as in the isotropic phase $f(\theta)= 1/2$ and $g(z)=d^{-1}$. 

The key non-dimensional input parameter in the theory is $\alpha$, with $0 \leq \alpha \leq 2$ \cite{McMillan1971}. The parameter $\alpha$ indicates the scale of the smectic ordering energy relative to that of the nematic ordering alone. We note that in this theory, high nematic ordering is required to induce smectic ordering, but the higher the value of $\alpha$, the lower the required degree of nematic order.  An extra term $\dsp -\frac{1}{2}\delta \tau^2$ which would favor spontaneous smectic ordering  in the absence of nematic ordering has been omitted (i.e $\delta=0$) \cite{McMillan1971,McMillan1972}. Such a term could be reintroduced, for example,  in a lamellar phase made from block copolymers, in which nematic order was induced by smectic order and not vice-versa, but we do not consider it here. 

The distribution functions $f(\theta)$ and $g(z)$ are now  determined self-consistently from the order parameters, angular functions and the smecticity parameter $\alpha$ \cite{KLZ1985}.  They can be compactly written in terms of non-dimensionalized potentials of mean torque, $U^*_\theta(\theta)$ and $U^*_z(z)$, as
\begin{subequations}
\be
f(\theta) = Q_\theta ^{-1} \exp \left[ -\frac{U^*_\theta(\theta)}{T^*} \right];
\ee
\be
g(z) = Q_z^{-1} \exp \left[ -\frac{ U^*_z(z)}{ T^*} \right].
\ee
\end{subequations}
The potentials of mean torque are given by
\begin{subequations}
\be
U^*_\theta(\theta) = -(1+\alpha \tau^2) S P_2(\cos{\theta});
\ee
\be
U^*_z(z) = -\alpha S^2 \tau \cos\left(\frac{2\pi z}{d}\right).
\ee
\end{subequations}
The partition functions $Q_\theta,Q_z$ are defined so as to normalize the distribution functions:
\begin{subequations}
\be
Q_\theta = \int_0^{\pi} \sin{\theta} d\theta \exp \left[ \frac{- U^*_\theta(\theta)}{T^*} \right];
\ee
\be
Q_z = \int_0^d dz \exp \left[ \frac{- U^*_z(z)}{ T^*} \right];
\ee
\be
Q=Q_{\theta}Q_z.
\ee
\end{subequations}

The equilibrium free energy can be rewritten as:
\be
\label{eq:KLZA1}
A^*= - T^* \log{Q} + (1/2)(S^2 + 3 \alpha \tau ^2 S^2),
\ee
which determines absolute phase stability if there is more than one equilibrium.\\

\begin{figure}[htb]
    \begin{center}
           \includegraphics[width=3.5in]{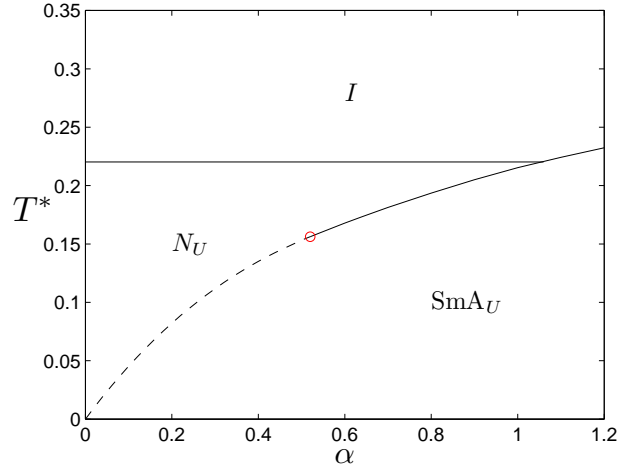}
    \end{center}
    \caption{Dependence of scaled transition temperature on smectic interaction parameter $\alpha$, following Kventsel \etal \cite{KLZ1985}. Continous lines: first order phase transitions;   broken lines: continuous transitions; circle:  $N_U-\Sm A_U$ tricritical point.}
\label{fig:KLZunidelta00}
\end{figure}

Fig. \ref{fig:KLZunidelta00} shows the phase diagram for this theory in the $(T^*, \alpha)$ plane. The theory predicts  $I$, $N_u$ and $\Sm A_u$ phases, with their relative stability tuned by varying  $\alpha$.  Smectic order is favored (i.e. the onset of smectic order occurs at higher temperatures) with increasing $\alpha$. For low $\alpha$ the $N_U-\Sm A_U$ transition is continuous, but it becomes first-order  at a tricritical point at $\alpha = 0.52$. At a critical value of $\alpha =\alpha_c \approx 1.02$, the $N$ phase disappears, and there is a direct $I-\Sm A_U$ transition. A number of features of the theory are not robust, such as the temperature at three-phase coexistence, and the independence of $T_{NI}$ of the smecticity parameter $\alpha$. But the main general features of the phase diagram are conserved in more detailed theories \cite, and seem to correspond to experiment.

\subsection{The biaxial nematic phase}
\label{subsec:biaxialbackground}
The molecular field theory for $D_{2h}$ biaxial nematics requires  a normalized distribution function $f(\Omega)$, and
four scalar order parameters ($S,D,P,C$) \cite{Straley1974, Rosso2007}, which  are orientational averages of  basis angular functions ($R_S,R_D,R_P,R_C$). We follow the notation convention of Teixeira \etal \cite{Teixeira2006}, with different scaling. The functions are:
\begin{subequations}
\be
R_S(\Omega)=P_2(\cos\theta);
\ee
\be
R_D(\Omega)=\sqrt{3/8} \sin^2{\theta} \cos 2{\phi};
\ee
\be
R_P(\Omega)=\sqrt{3/8} \sin^2{\theta} \cos 2{\psi};
\ee
\begin{eqnarray}
R_C(\Omega)=(1/2)\left(1+\cos^2{\theta} \right) \cos{2\phi} \cos{2\psi}\nonumber \\
-\cos{\theta} \sin {2\phi} \sin{2\psi}.
\end{eqnarray}
\end{subequations}
The order parameters are defined in terms of  basis angular function averages: 
\be
\label{Dopssmectic}
i = \langle R_i \rangle = \int_\Omega  d\Omega f(\Omega) R_i(\Omega),
\ee
with $i\in\{S,D,P,C\}$. We note that (subject to exchange of axes) in the $N_U$ phase, in general, $S,D \ne 0$, but $C=P=0$. In biaxial phases all four order parameters in general are non-zero. 

The degree of biaxiality is controlled by two parameters, ($\gamma,\lambda$) (see e.g. \cite{Virga2003}), which enter the non-dimensional internal energy as follows: 
\begin{eqnarray}
\label{eq:nemainterenergy}
U^*_N = -\frac{1}{2}  \left(S ^2 + 4\gamma S D + 4\lambda D ^2 \right. \nonumber \\
\left. + 2 ( P ^2 + 2\gamma P C + \lambda C ^2) \right). 
\end{eqnarray}
The case $\lambda=\gamma=0$ corresponds to uniaxial $D_{\infty h}$ molecules.  
The  orientational distribution function $f(\Omega)$  is derived by minimizing a Helmholtz free energy functional:
\begin{equation}
\label{eq:nematfreeenergy}
A^*= -T^*\int_{\Omega} d\Omega f(\Omega)\ln\left(8\pi^2f(\Omega)\right) +U^*_N . 
\end{equation}

The equilibrium orientational distribution function $f(\Omega)$ is expressed in term of the potential of mean torque
\begin{eqnarray}
\label{eq:pmtNb}
U^*_\Omega (\Omega) =-\Big[ (S + 2 \gamma D) R_S(\Omega) + 2(\gamma S + 2\lambda D) R_D(\Omega) \nonumber\\
+ 2 (P + \gamma C) R_P(\Omega) 
+ 2 (\gamma P + \lambda C) R_C(\Omega) \Big], \nonumber \\
\end{eqnarray}
and is given by
\be
f(\Omega) = Q_\Omega^{-1} \exp \left[ -\frac{U^*_\Omega (\Omega)}{T^*} \right],
\ee
where the partition function $Q_\Omega$ is defined in order to normalise $f(\Omega)$:
\be
Q_\Omega = \int_\Omega d\Omega \exp \left[ -\frac{U^*_\Omega (\Omega)}{T^*} \right].
\ee
The phase stability of the system is determined from the equilibrium Helmholtz free energy:
\begin{eqnarray}
\label{eq:HelmholtzfeNb}
A^* =  - T^* \log Q_\Omega + \frac{1}{2} \Big( S ^2 + 4\gamma S D + 4\lambda D ^2 \nonumber \\
+ 2 ( P ^2 + 2\gamma P C + \lambda C ^2) \Big).
\end{eqnarray}

In general both $\gamma,\lambda$ are required to define the full biaxiality. There are two particular limits which enable the biaxiality to be expressed in terms of a single parameter. We now discuss these in turn. 

\subsubsection{The geometric mean approximation (GMA)} 
\label{sec:GMAnematic}
\begin{figure*}[ht]
    \begin{center}
    		\subfloat[Geometric mean (GMA)]{
            \includegraphics[width=0.35\textwidth]{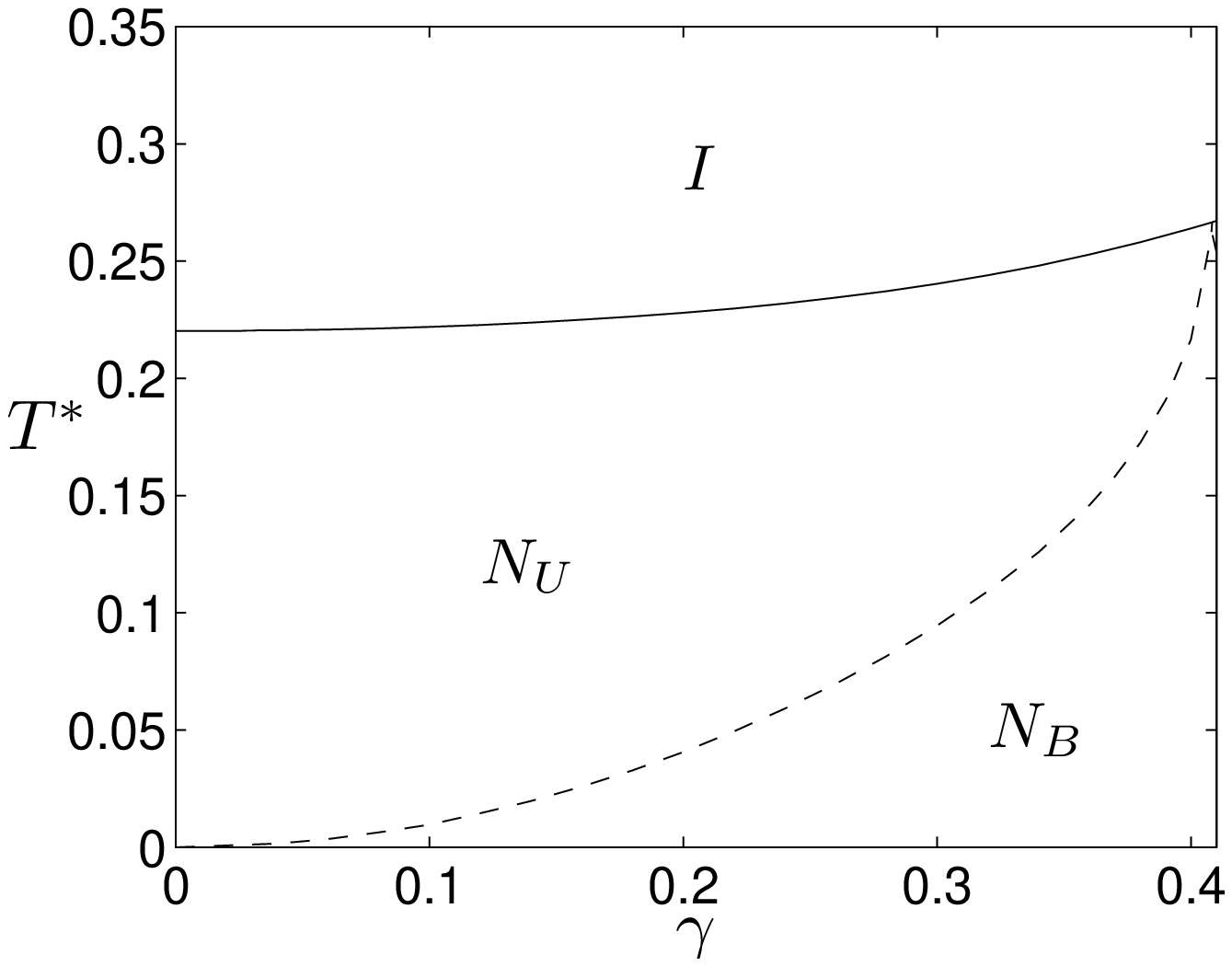}
		\label{fig:GMphasemap}
        }
        \subfloat[Sonnet-Virga-Durand (SVD)]{
            \includegraphics[width=0.35\textwidth]{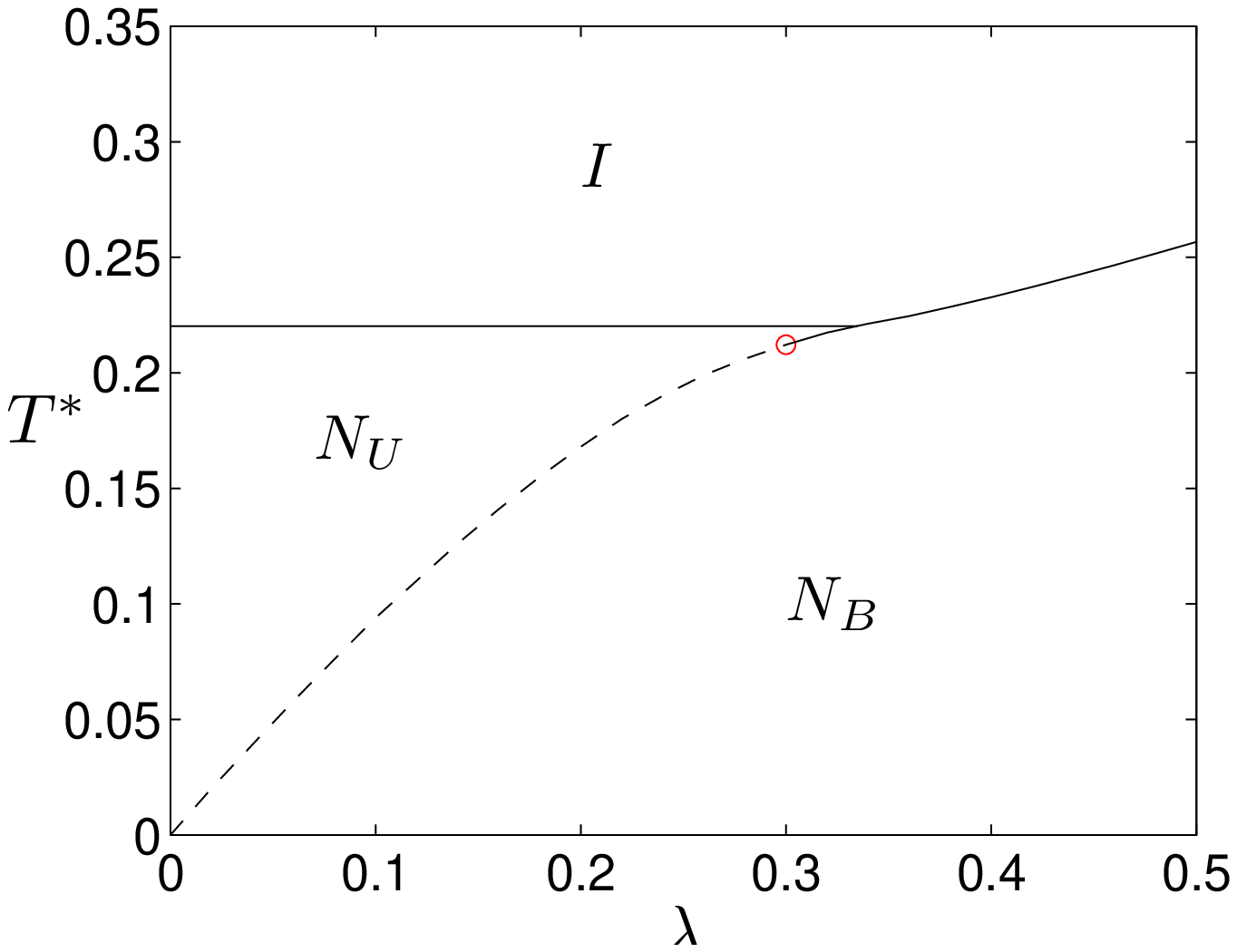}
		\label{fig:SVDphasemap}
        }
    \end{center}
    \caption{
        Phase diagrams in biaxiality-temperature plane in the geometric mean (GMA) (a) and Sonnet-Virga-Durand (SVD) (b) approximations. Continous lines: first order phase transitions;  broken lines: continuous transitions;  red circle: $N_U-N_B$ tricritical point.
     }
\label{fig:GMAvsSVD}
\end{figure*}

The GMA  limit   $\gamma=\lambda^2$ emerges naturally from a consideration of dispersive forces or of geometric shapes \cite{Straley1974,Luckhurst1975,Boccara1976,RemlerHaymet1986,Virga2003}. For  a long time was regarded as generic.  The interesting parameter range is  $\dsp 0 \le \gamma \le  \gamma_c=\frac{1}{\sqrt{6}} \approx 0.408$.   At $\gamma_c$, the biaxiality is in some sense maximal, and the roles of major and minor axes are exchanged, so that values of $\gamma>\gamma_c$ can be mapped onto values of $\gamma < \gamma_c$.  Further discussion can be found e.g. in ref.[\onlinecite{Virga2003}].  The key feature of the resulting phase diagram is that there is a first-order $I-N_U$ phase transition, followed at a lower temperature by a continuous $N_U-N_B$ transition. These transitions collapse onto a single multicritical point (the Landau point) at $\gamma=\gamma_c$. 

In the GMA,   it can be shown, using eqs.(\ref{eq:pmtNb},\ref{eq:HelmholtzfeNb}), that the analysis requires only two order parameters, rather than the full complement of four. These  are defined as follows:
\begin{subequations}
\be
\label{eq:J1SDPC}
{\cal J}_1 = \langle J_1(\Omega) \rangle = \langle R_S(\Omega) + 2\gamma R_D(\Omega) \rangle = S + 2\gamma D,
\ee
\be
\label{eq:J2SDPC}
{\cal J}_2 = \langle J_2(\Omega) \rangle = \langle R_P(\Omega) + \gamma R_C(\Omega) \rangle = P + \gamma C.
\ee
\end{subequations}
In the $I$ phase  both order parameters are zero, in the $N_U$ phase,  $\meanJ_1  \ne 0$, but $\meanJ_2=0$, while in the $N_B$ phase neither order parameter is zero.

\subsubsection{The Sonnet-Virga-Durand approximation (SVD)}
\label{sec:SVDnematic}
In this limit \cite{Virga2003, Virga2007stability} $\gamma=0$.  The relevant range is now  $\dsp 0 \leq \lambda \leq \frac{1}{2}$.  In the SVD the so-called minor order parameters $D$ and $P$ identically vanish.  Hence again there are only two order parameters, $S,C$, with $C=0$ in the $N_U$ phase.   

There are two key differences between GMA and SVD. Firstly, in  GMA, there is no direct $I-N_B$ transition except at a continuous transition at the single point of maximum biaxiality, usually known as the Landau multicritical point, where $I,N_U,N_B$ coexist.   By contrast,  SVD predicts the whole line of first-order $N_B-I$ transitions. Secondly, GMA predicts that the $N_B-N_U$ transition is always continuous, whereas  in  SVD,  the $N_B-N_U$ transition is continuous for a long range of biaxiality and first order for a small range of biaxiality, with a tricritical point on the $N_B-N_U$ transition line.  Apart from this difference, the two phase diagrams are topologically similar, as can be seen in Figs.\ref{fig:GMAvsSVD}, where we show the phase diagrams for GMA and SVD as functions of  biaxiality  and temperature.

\section{Molecular field theory of biaxial smectic phases.}
\label{sec:MFTsection}
\setcounter{subsubsection}{0}

\subsection{Background} 
\label{subsec:background}
A formal derivation of the molecular field theory for biaxial smectic A phases, formed of molecules  with $D_{2h}$ symmetry, has been given by Teixeira \textit{et al.} \cite{Teixeira2006}. This is a combination of the molecular field theory for biaxial nematics and the McMillan theory for uniaxial smectic A. In addition, Teixeira \textit{et al.} \cite{Teixeira2006} also discussed an alternative model based on the approximation by Kventsel, Luckhurst and Zewdie (KLZ) \cite{KLZ1985}. In this section we discuss relevant equations for our calculation.  

\subsection{Distribution function and order parameters} 
\label{subsec:smop}
As in section \ref{subsec:smecticbackground},   the distribution function $F(z,\Omega)$ will be decoupled using the KLZ approximation \cite{KLZ1985}, yielding $F(z,\Omega)=f(\Omega)g(z)$, with $\Omega=(\psi,\theta,\phi)$ as in sec. \ref{subsec:biaxialbackground}.  

 The principal order parameters are:
\begin{itemize}
\item[(a)]  The biaxial  nematic order parameters $i=\langle R_i\rangle$, for $i\in \{S,P,C,D\}$,  as defined in  eq.\eqref{Dopssmectic}.  As in section \ref{subsec:biaxialbackground}, we  shall consider GMA and SVD flavors, which both reduce this order parameter set to two independent parameters, one representing uniaxial order and one biaxial order.  
\item[(b)] The simple smectic order parameter 
\begin{align}
\label{eq:opsmectic1}
&\tau=\left\langle \cos \left(\frac{2\pi z}{d}\right) \right\rangle \nonumber \\
&= \int_{\Omega}d\Omega \int_0^d dz F(z,\Omega)\cos \left(\frac{2\pi z}{d}\right) \nonumber \\
&=\int_0^ d dz g(z)\cos\left(\frac{2\pi z}{d}\right)
\end{align}
as in eq.\eqref{eq:op4}, representing the magnitude of the density modulation.  Note that we have used the decoupling approximation to eliminate the angular integral. 
\item[(c)] The amplitudes of the periodic modulations the nematic order parameters, combining eqs.(\ref{eq:op3},\ref{Dopssmectic}):
\begin{align}
\label{eq:sigmaopssmectic}
&\sigma _i = \left\langle \cos{\left(\frac{2\pi z}{d}\right)} R_i(\Omega) \right\rangle \nonumber \\
&= \int_{\Omega} d\Omega \int_0^d  dz F(z,\Omega)\cos{\left(\frac{2\pi z}{d}\right)} R_i(\Omega) \nonumber \\
&=\int_{\Omega}d\Omega f(\Omega)R_i(\Omega)
\int_0^d  dz g(z)\cos{\left(\frac{2\pi z}{d}\right)} \nonumber \\
&= \langle R_i\rangle \tau,
\end{align}
for $i \in\{S,P,C,D\}$.
\end{itemize} 

\subsection{Internal energy}
\label{subsec:internegy}
The internal energy of the system consists of two parts:  $U^*_N$ and $U^*_S$, which can be associated respectively with purely nematic energy and  a smectic contribution coming from orientational-translational coupling: 
\begin{equation}
\label{eq:internegy1}
U^* = U^*_N + U^*_S.
\end{equation}
The nematic free energy  is  as given in eq.\eqref{eq:nemainterenergy}
\be
\label{eq:nemaenergy}
U^*_N = -\frac{1}{2} \Big(S ^2 + 4\gamma S D + 4\lambda D ^2 + 2 ( P ^2 + 2\gamma P C + \lambda C ^2) \Big). 
\ee

The second part of the internal energy  is the mixed orientational-translational interaction,  which  we write  by analogy with eqs.(\ref{eq:helmholtz1},\ref{eq:nemainterenergy}) as:
\begin{eqnarray}
\label{eq:internegy2}
U^*_S = -\frac{1}{2}  \alpha \Big( \sigma_S ^2 + 4\gamma' \sigma_S \sigma_D + 4\lambda' \sigma_D ^2 \nonumber \\
+  2 ( \sigma_P ^2 + 2\gamma' \sigma_P\sigma_C + \lambda' \sigma_C ^2) \Big). 
\end{eqnarray}
In connection with the smectic contribution, we note that:
\begin{itemize}
\item[(a)]  it is weighted by the  same factor $\alpha$ (and for the same reason)   as occurs in the McMillan theory eq.\eqref{eq:helmholtz1}; 
\item[(b)] the molecules possess $D_{2h}$ symmetry, and so the degree of biaxiality must be characterized by parameters $\gamma',\lambda'$ which bear the same relationship to the smectic potential interaction as $\gamma,\lambda$ do to the nematic biaxial interaction.  
\end{itemize}
In order to simplify eq.\eqref{eq:internegy2}:
\begin{itemize}
 \item[(i)] we recall the distribution function decoupling result  $\sigma_i=\langle R_i\rangle\tau$.  
\item[(ii)] we suppose $\gamma'=\gamma;\;\lambda'=\lambda$, yielding the following result for the mixed orientational-translational interaction energy: 
\begin{eqnarray}
\label{eq:internegy3}
U^*_S = -\frac{1}{2}  \alpha \tau^2  \Big(S ^2 + 4\gamma S D + 4\lambda D ^2 \nonumber \\
+ 2 ( P ^2 + 2\gamma P C + \lambda C ^2) \Big).    
\end{eqnarray}
\end{itemize}

Finally, we note that in principle,  a third contribution of form $U^*_0 =  -\frac{1}{2} \delta \tau ^2$ is possible. We omit this term, partly for simplicity,  partly because it refers to spontaneous layer formation, not driven by the orientational potential, and partly because we are concentrating on predicting trends. As discussed  in sec.\ref{subsec:smecticbackground}, we shall neglect this term (i.e. $\delta=0$).  For detailed agreement with experimental results (e.g. see ref.[\onlinecite{McMillan1972}]), it may be necessary to include it.   Our final expression for the total internal energy is thus:
\begin{align}
\label{eq:internegy4}
&U^*= -\frac{1}{2} \left(1+\alpha\tau^2\right)\Big(S ^2 + 4\gamma S D + 4\lambda D ^2 
\nonumber \\
&+ 2 ( P ^2 + 2\gamma P C + \lambda C ^2) \Big) =-\left(1+\alpha\tau^2\right)U^*_N.
\end{align}

\subsection{Helmholtz free energy and potential of mean torque}
\label{subsec:freeenergy}
The  liquid crystalline contribution to Helmholtz free energy, per particle, is given by
\begin{align}
\label{eq:helmholtz2}
&A^*= T^*\int_\Omega d\Omega f(\Omega)\ln\Big(8 \pi ^2 f(\Omega)\Big) \nonumber \\
&+ T^*\int_0^d dz g(z)\ln\Big(dg(z)\Big) + \left(1+\alpha\tau^2\right)U^*_N.
\end{align}
We note that in eq.\eqref{eq:helmholtz2}, as is usual in liquid crystal calculations, the isotropic phase contribution to the Helmholtz free energy has been subtracted.  Thus in the isotropic limit, in which $\dsp f(\Omega)=\frac{1}{8\pi^2};\; g(z)=\frac{1}{d}$, $A^*=0$. 

The orientational and translational distribution functions can be determined by minimizing the free energy given in eq.\eqref{eq:helmholtz2} and can be written compactly in terms of the potentials of mean torque as
\begin{subequations}
\be
f(\Omega) = Q_\Omega ^{-1} \exp \left[ -\frac{U^*_\Omega(\Omega)}{T^*} \right];
\ee
\be
g(z) = Q_z^{-1} \exp \left[ -\frac{ U^*_z(z)}{ T^*} \right],
\ee
\end{subequations}
where the potentials of mean torque are given by
\begin{align}
\label{eq:orientpairpot}
&U_\Omega^* (\Omega) = - (1+\alpha \tau ^2) \Big[ (S + 2 \gamma D) R_S(\Omega) \nonumber \\
&+ 2(\gamma S + 2\lambda D) R_D(\Omega)) + 2 (P + \gamma C) R_P(\Omega) \nonumber \\
&+ 2 (\gamma P + \lambda C) R_C(\Omega) \Big]; 
\end{align}
\be
U_z^* (z) = - 2 \alpha \tau U_N^* \cos{(2 \pi z/d)}.
\ee
The partition functions $Q_\Omega,Q_z$ are defined so as to normalize the distribution functions:
\begin{subequations}
\be
Q_\Omega = \int_\Omega d\Omega \exp \left[ \frac{- U^*_\Omega(\Omega)}{T^*} \right];
\ee
\be
Q_z = \int_0^d dz \exp \left[ \frac{- U^*_z(z)}{ T^*} \right];
\ee
\be
Q= Q_{\Omega}Q_z.
\ee
\end{subequations}
At a given temperature, the value of the free energy at an equilibrium point is given by
\begin{equation}
\label{eq:febiaxsmec}
A^* = - T^* \log{Q} + (1+3 \alpha \tau ^2)U^*_N.
\end{equation}

In subsections \ref{sec:GMAnematic} and \ref{sec:SVDnematic}, we have discussed the geometric mean (GMA) and Sonnet-Virga-Durand (SVD) approximations, which we now use in our biaxial smectic A phase calculation. We recall that these approximations enable the internal energy (here eq. \eqref{eq:nemaenergy}), and the orientational potential of mean torque  (here eq. \eqref{eq:orientpairpot}) to be rewritten compactly in terms of  only two order parameters,  controlled by a single biaxiality parameter.

\begin{itemize}
\item[(a)] \textbf{GMA:}
The two independent composite orientational order parameters have been given in eqs. (\ref{eq:J1SDPC},\ref{eq:J2SDPC}),  and are  ${\cal J}_1$ (non-zero in  all liquid crystal phases) and  ${\cal J}_2$ (only non-zero in biaxial phases). 
The nematic internal energy and potential of mean torque in eqs. \eqref{eq:nemaenergy} and \eqref{eq:orientpairpot} can be rewritten compactly in terms of ${\cal J}_1,{\cal J}_2$
\be
U_N^* = - \frac{1}{2}({\cal J}_1^2 + 2{\cal J}_2^2);
\ee
\be
U^*_\Omega (\Omega) = -(1+\alpha \tau^2)\Big({\cal J}_1J_1(\Omega) + 2{\cal J}_2J_2(\Omega) \Big).
\ee
Thus in this model, there are three self-consistency equations, given by eq. \eqref{eq:opsmectic1} and
\be
\label{eq:KLZGMAselfcon}
{\cal J}_i = \int_\Omega d\Omega f(\Omega) J_i(\Omega),
\ee
with $i\in\{1,2\}$.

\item[(b)]\textbf{SVD:}
 The two orientational order parameters  are $S$ (non-zero in  all liquid crystal phases) and $C$ (only non-zero in biaxial phases. The nematic internal energy and potential of mean torque in eqs. \eqref{eq:nemaenergy} and \eqref{eq:orientpairpot} can be rewritten compactly in terms of $S,C$
\be
U_N^* = - \frac{1}{2}(S^2 + 2\lambda C^2);
\ee
\be
U^*_\Omega (\Omega) = -(1+\alpha \tau^2)\Big(S R_S(\Omega) + 2\lambda C R_C(\Omega)\Big).
\ee
Thus in this model, there are three self-consistency equations, given by eq. \eqref{eq:opsmectic1} and
\be
\label{eq:KLZSVDselfcon}
i = \langle R_i \rangle = \int_\Omega d\Omega f(\Omega) R_i(\Omega),
\ee
with $i\in\{S,C\}$.
\end{itemize}

\subsection{Method}
The procedure  to determine the stable state at a given temperature involves solving the relevant self-consistent equations for both the GMA and the SVD models. In each case we solve the three self-consistency equations, eqs.(\ref{eq:opsmectic1}, \ref{eq:KLZGMAselfcon}) for GMA and (\ref{eq:opsmectic1},\ref{eq:KLZSVDselfcon}) for SVD, using the MATLAB function \textit{fsolve}. This function uses an improved Newton's method (the so-called  trust-region dogleg algorithm) and iterates towards a solution from a given starting point. The chosen solution is the one that give the lowest value for the free energy in eq. \eqref{eq:febiaxsmec}. A phase transition occurs at the point at which one or more order parameters change from a zero to a non-zero value as the temperature is lowered.

\section{Numerical results}
\label{sec:results}
We  present results from both the GMA and SVD flavors of our theory, and then discuss what features these two sets of results possess in common. 

\subsection{GMA}
\label{sec:KLZGMAresults}

\begin{figure*}[ht]
    \begin{center}
        \subfloat[$\alpha = 0$]{
            \includegraphics[width=0.35\textwidth]{GMphasemap}
		\label{fig:GMphasemap1}
        }
        \subfloat[$\alpha = 0.3$]{
           \includegraphics[width=0.35\textwidth]{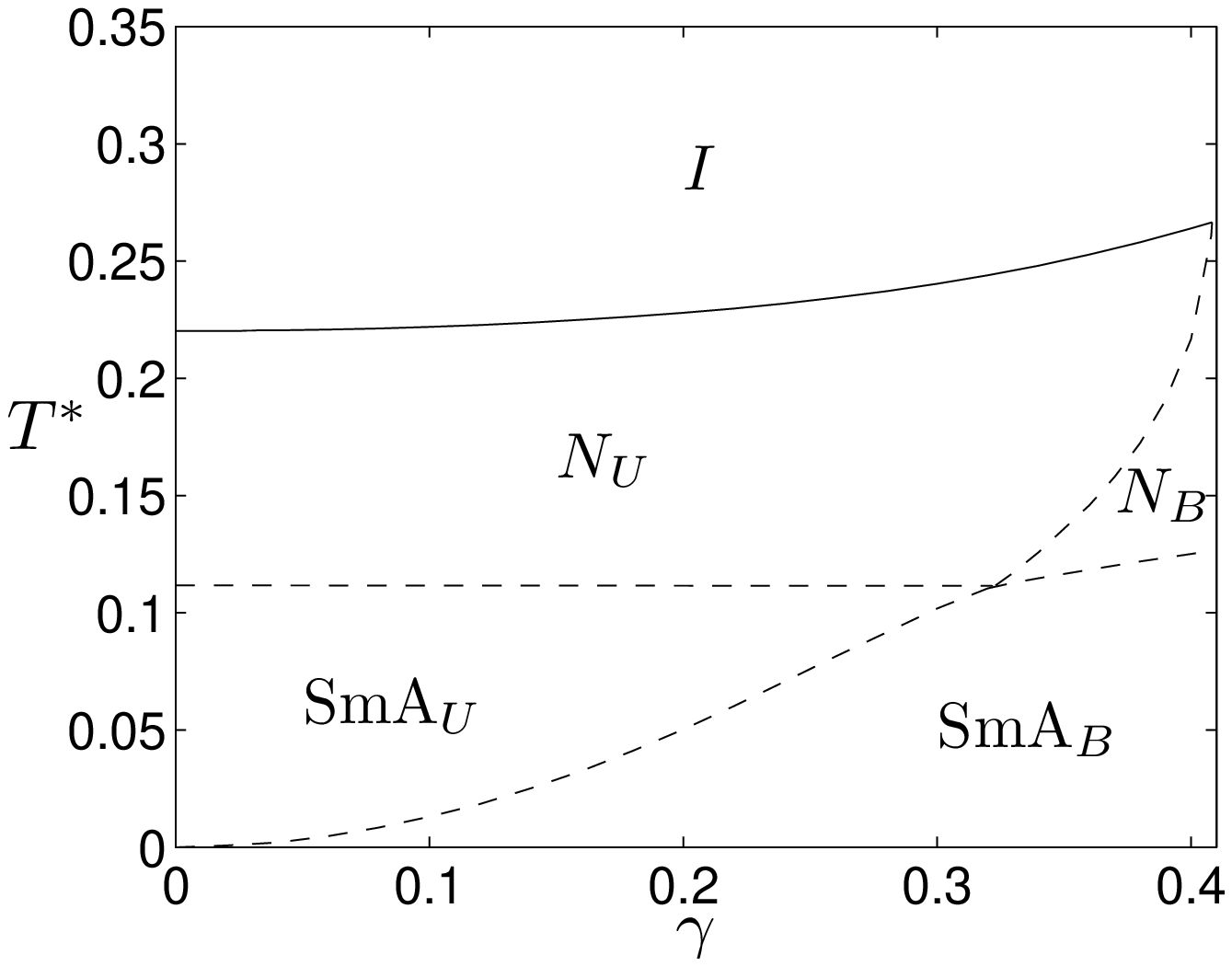}
           \label{fig:KLZGMphasemapdelta00alpha03}
        }\\ 
        \subfloat[$\alpha = 0.9$]{
            \includegraphics[width=0.35\textwidth]{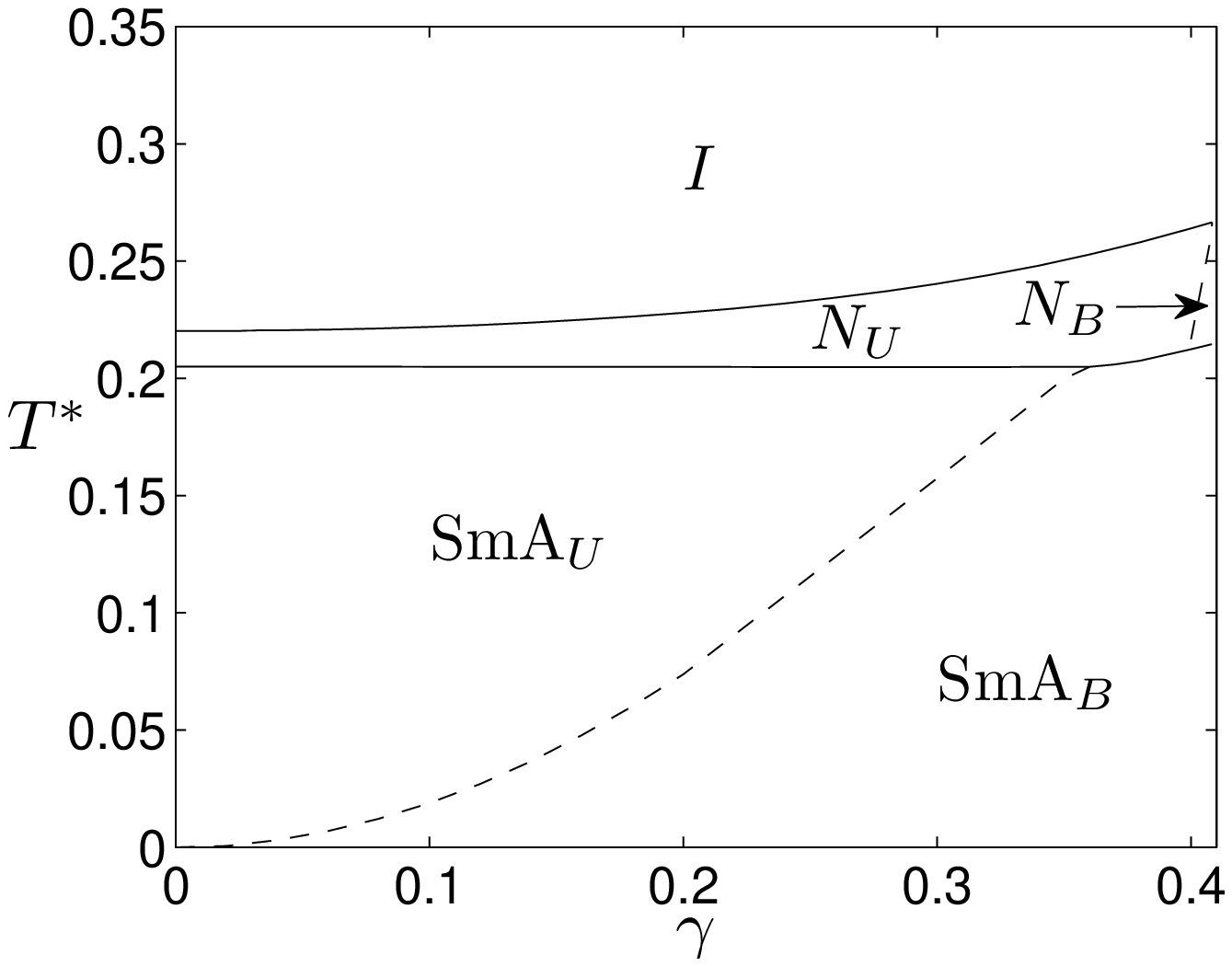}
            \label{fig:KLZGMphasemapdelta00alpha09}
        }
        \subfloat[$\alpha = 1.2$]{
           \includegraphics[width=0.35\textwidth]{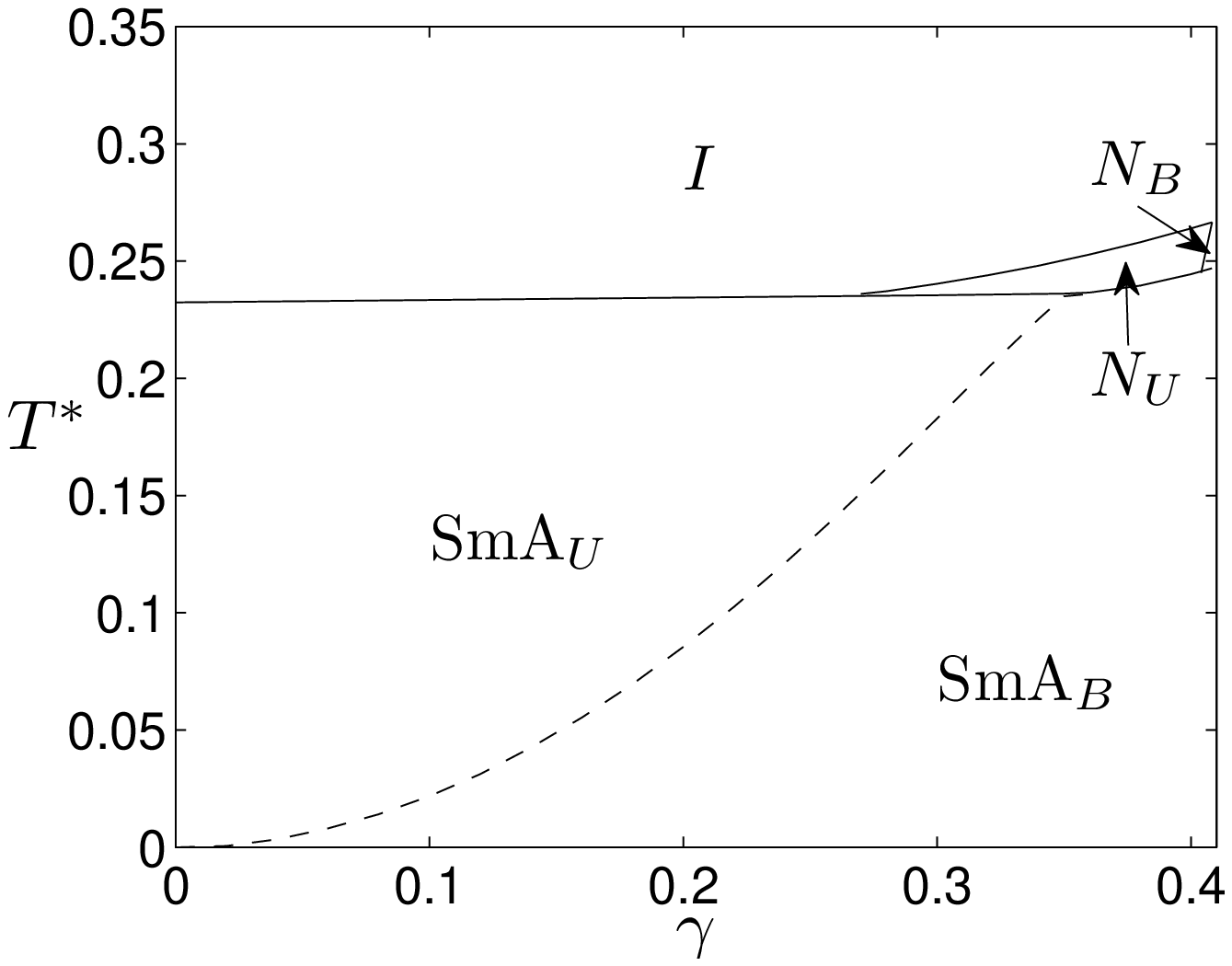}
           \label{fig:KLZGMphasemapdelta00alpha12}
        }\\
	\subfloat[$\alpha = 1.5$]{
            \includegraphics[width=0.35\textwidth]{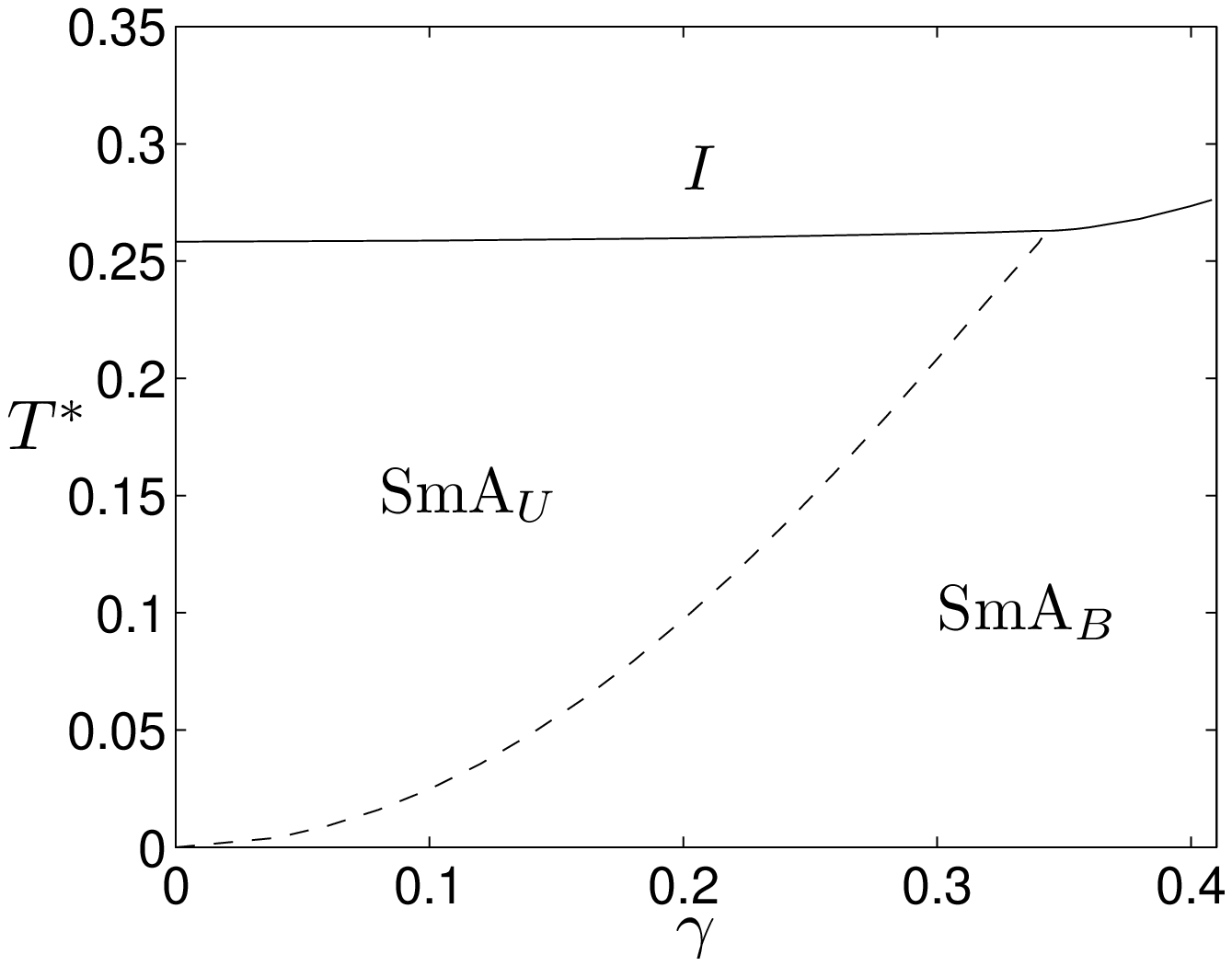}
            \label{fig:KLZGMphasemapdelta00alpha15}
        }
    \end{center}
    \caption{Geometric mean approximation phase diagrams  as a function of biaxiality $\gamma$ and scaled temperature $T^*$, for representative values of the McMillan smectic parameter $\alpha$. Continuous lines: first order phase transitions;  broken lines: continuous transitions. }
\label{fig:KLZGMphasemapsdelta00}
\end{figure*}

In Figs.\ref{fig:KLZGMphasemapsdelta00} we show a representative set of five  GMA phase diagrams for increasing $\alpha$. Each phase diagram shows a constant $\alpha$ slice in the  $(\alpha-\gamma-T^*)$ space.  All topological configurations of phases that we find are shown in one of these examples. We recall that in zero smecticity case ($\alpha=0$) the stability of the biaxial nematic phase $N_B$  increases relative that of the uniaxial nematic $N_U$ on increasing $\gamma$,  up to the Landau multicritical point at $\dsp \gamma = \gamma_c= \frac{1}{\sqrt{6}}$.  Symmetry around $\gamma_c$ then implies that  the same series of phases occurs  for $\gamma \ge \gamma_c$ as for $\gamma \le \gamma_c$. We expect this symmetry also to apply for non-zero $\alpha$, and our pictures only include the cases $\gamma \le \gamma_c$.  

For completeness, we include in  Figs.\ref{fig:KLZGMphasemapsdelta00}  the zero smecticity case $\alpha = 0$ phase diagram, already discussed   in section \ref{subsec:biaxialbackground} (see Fig.\ref{fig:GMphasemap}).  Here, except at $\gamma =0$, the ground state is always $N_B$. For low $\gamma$, the $N_B$ phase only intercedes at very low temperatures. Only for $\gamma$ very close to $\gamma_c$ does the $N_U-N_B$ phase boundary noticeably increase in temperature. Then  at the Landau multicritical point  $\gamma=\gamma_c$, all phases  coincide.    

For $\alpha =0.3$, shown in Fig.\ref{fig:KLZGMphasemapdelta00alpha03}, the higher temperature part of this diagram ($T^* \agt 0.1$) is unchanged from  the  $\alpha=0$ case, indicating that the orientational-translational coupling is sufficiently low that low temperatures are required to activate the smectic modulation.   But now at sufficiently low temperatures both uniaxial $\mathrm{SmA}_U$ and biaxial smectic $\mathrm{SmA}_B$ phases are stabilized, with the stability of the $\mathrm{SmA}_B$ phase increasing significantly at larger $\gamma$. 

The ground state $N_B$ phase is unsurprisingly preempted by a $\Sm A_B$ phase, while for values of  $\gamma$ for which the $N_B$ phase requires very low temperatures, the $N_U$ phase yields to a $\Sm A_U$ phase.  This picture seems generic for low positive $\alpha$.   Although the $N_U$  and even $N_B$ phases are retained at higher temperatures,  the $N_B$ phase is now confined to a window around  the Landau multicritical point. In the $\alpha=0.3$ case, this corresponds to  $T^* \agt 0.5 T_{NI}$, and $0.32 \alt \gamma \alt 0.41$,   The temperature at  the $N_U-\Sm A_U$ phase boundary is independent of  $\alpha$, and the transition is continuous. Likewise the $N_U-N_B$ transition remains continuous at finite $\alpha$, as are the new $N_B-\Sm A_B$ and $\Sm A_u-\Sm A_B$ transitions.   An interesting, if implausible,  stable feature of this phase diagram is a four-phase coexistence point at $\gamma \approx 0.32, T^* \approx 0.11$, at which  the three continuous phase transition lines $N_U-N_B,\;N_U-\Sm A_U,\;\Sm A_U-\Sm A_B$ collide.

For   $\alpha=0.9$  (see Fig.\ref{fig:KLZGMphasemapdelta00alpha09}), the stability region of the nematic phases is much smaller. In addition, the $\mathrm{SmA}_U-N_U$ transition is  now first order, consistent with  the $\gamma = 0$ phase diagram in Fig.\ref{fig:KLZunidelta00}.  As in the $\alpha=0.3$ case, the $\mathrm{SmA}_B$ phase is increasingly stable at higher $\gamma$.  The $\mathrm{SmA}_B-\mathrm{SmA}_U$ transition is still second order, but now the $\mathrm{SmA}_B-N_B$ and $\mathrm{SmA}_B-N_U$ transitions are first order. We note also that the $\alpha=0.3$  four-phase $N_U,N_B,\Sm A_U,\Sm A_B$ coexistence point has split into two separate critical end points, one for $N_U,\Sm A_U,\Sm A_B$ and one for   $N_U,N_B,\Sm A_B$. The already narrow $N_B$ stability has shrunk further, and is now just a sliver for $ 0.4 \alt \gamma \alt 0.41$, and (at its greatest) $T^*/T^*_{NI} \agt 0.85$.  

For $\alpha = 1.2$ (see Fig.\ref{fig:KLZGMphasemapdelta00alpha12}), the $N_U$ phase is pre-empted by the $\Sm A_U$ phase over most of its range, and the $I-N_U$ transition line is replaced by  a line of direct $I-\Sm A_U$  transitions.  This feature of the phase diagram is consistent with the smectic molecular field theory  \cite{KLZ1985}, shown in Fig. \ref{fig:KLZunidelta00} for $\gamma = 0$. The nematic phase region is now confined to a  narrow temperature range close to the Landau multicritical point; $N_B$ phase is just hanging on in a tiny region in the immediate neighborhood of  the Landau multicritical point.  

In the the final case we consider, $\alpha=1.5$ (see Fig. \ref{fig:KLZGMphasemapdelta00alpha15}), the orientational-translational coupling is now large enough to induce smectic order as soon as nematic order appears.  Now the narrow regions of  $N_U,N_B$ stability have both been overtaken by smectic order. For $\gamma \alt 0.34$,  there is a direct $\mathrm{SmA}_U-I$ phase transition, while  for $0.34 \alt \gamma \alt \gamma_c \approx 0.41$, there is a direct $\mathrm{SmA}_B-I$ phase transition line.  At lower temperatures for all $\gamma \alt 0.34$, there is a further continuous $\Sm A_U -\Sm A_B$ transition.   

\begin{figure}[htb]
    \begin{center}
           \includegraphics[width=3.5in]{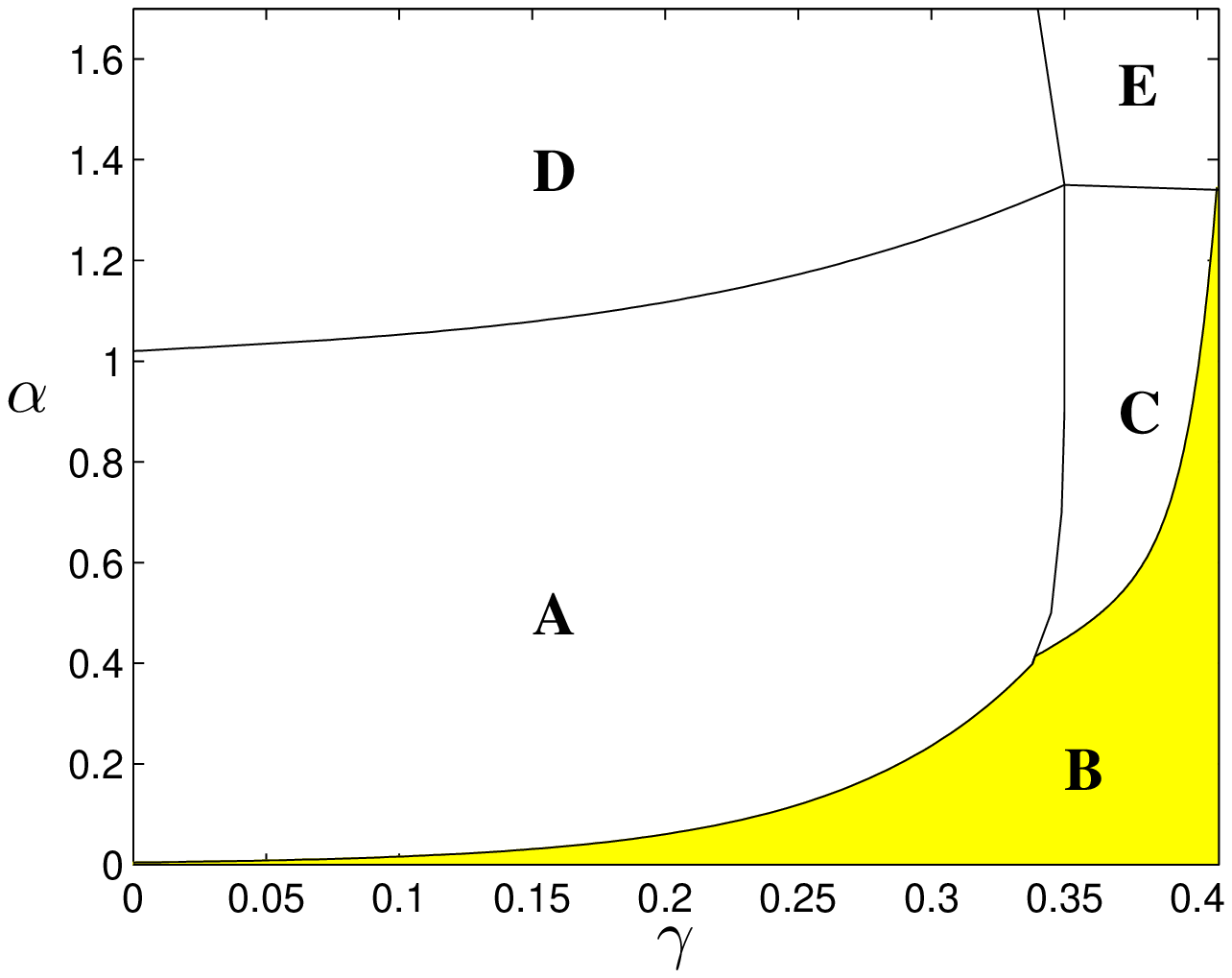}
    \end{center}
    \caption{GMA: phase sequences observed (approximate) in the $(\gamma,\alpha)$ (biaxiality-smecticity) plane.  The region with stabilized $N_B$ is shaded (yellow online). \newline 
		\begin{tabular}{llllllll}
		\textbf{A}:& $\Sm A_B-\Sm A_U-N_U-I~~~$&&  \textbf{B}:&$\Sm A_B-N_B-N_U-I$ \\
		\textbf{C}:& $\Sm A_B-N_U-I~~~$ &&  \textbf{D}:& $\Sm A_B-\Sm A_U-I$ \\
		\textbf{E}:&$\Sm A_B-I$&&
		\end{tabular}
		}
\label{fig:GMABiaxVSsmec}
\end{figure}

Finally in  Fig.\ref{fig:GMABiaxVSsmec}, we summarize these results, showing  the phase sequences which occur, as a function of the two control parameters  $\gamma, \alpha$.  We postpone a detailed discussion of this diagram. However, we can note immediately there are five predicted phase sequences. But  only one  out of these five -- the shaded region \textbf{B}  in  Fig.\ref{fig:GMABiaxVSsmec} --  includes the biaxial nematic $N_B$. and this is a rather restricted region of the control parameter plane.   

\subsection{SVD}
\label{sec:KLZSVDresults}

\begin{figure*}[ht]
    \begin{center}
        \subfloat[$\alpha = 0$]{
            \includegraphics[width=0.35\textwidth]{SVDphasemap}
		\label{fig:SVDphasemap1}
        }
        \subfloat[$\alpha = 0.3$]{
           \includegraphics[width=0.35\textwidth]{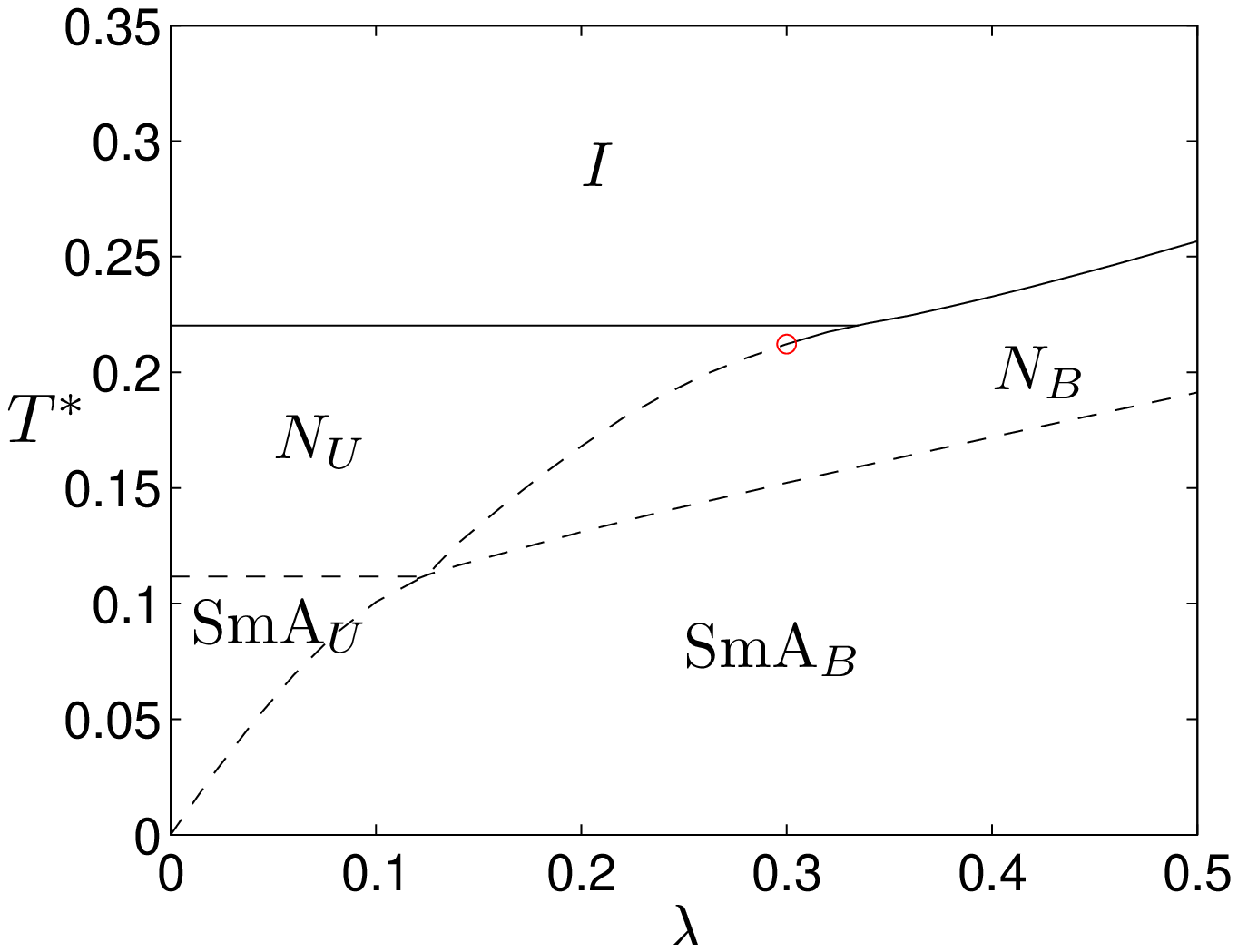}
           \label{fig:KLZSVDphasemapdelta00alpha03}
        }\\ 
        \subfloat[$\alpha = 0.9$]{
            \includegraphics[width=0.35\textwidth]{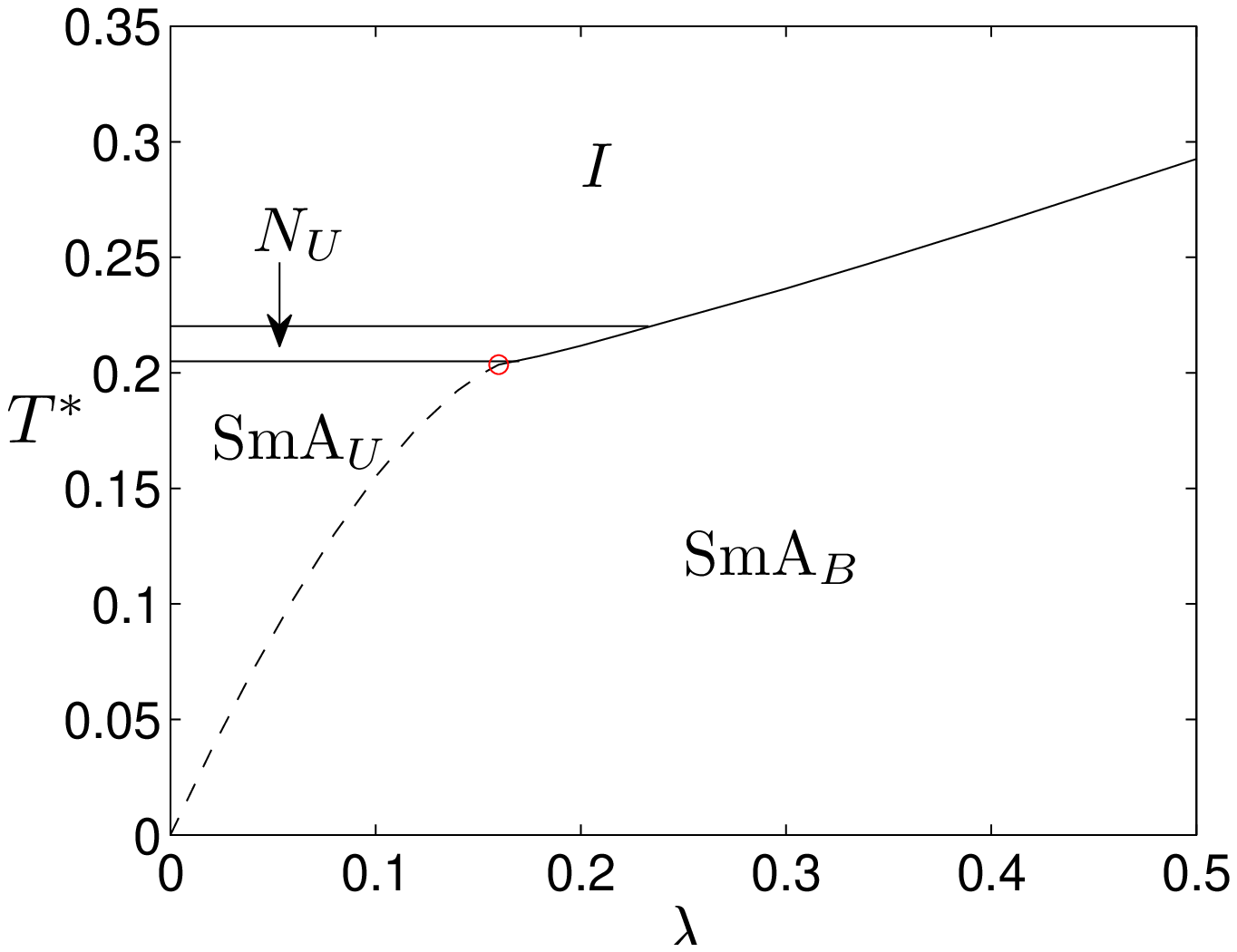}
            \label{fig:KLZSVDphasemapdelta00alpha09}
        }
        \subfloat[$\alpha = 1.2$]{
           \includegraphics[width=0.35\textwidth]{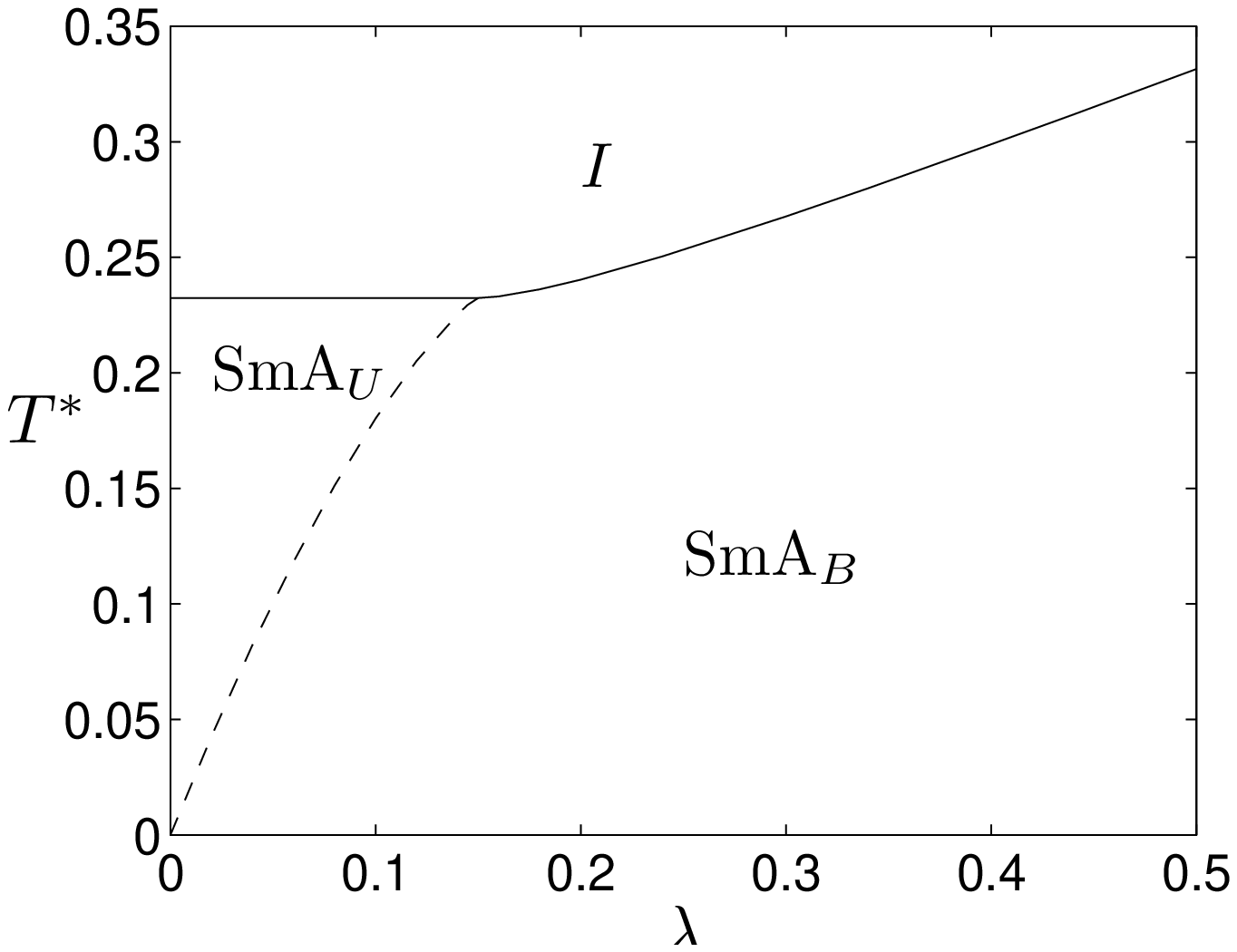}
           \label{fig:KLZSVDphasemapdelta00alpha12}
        }
    \end{center}
    \caption{
      SVD approximation phase diagrams  as a function of biaxiality $\gamma$ and scaled temperature $T^*$, for representative values of the McMillan smectic parameter $\alpha$. Continous lines: first order phase transitions;  broken lines: continuous transitions; red circle: tricritical point.}
\label{fig:KLZSVDphasemapsdelta00}
\end{figure*}

In Figs.\ref{fig:KLZSVDphasemapsdelta00} we show a representative set of four SVD phase diagrams for increasing $\alpha$. Each phase diagram shows a constant $\alpha$ slice in the $(\alpha-\lambda-T^*)$ space. As in the GMA case, all topological configurations of phases that we find are shown in one of these examples. We recall that in zero smecticity case ($\alpha = 0$) the biaxiality parameter is restricted to $0 \leq \lambda \leq 0.5$.

For completeness, we include in Fig.\ref{fig:SVDphasemap1} the zero smecticity case $\alpha = 0$ phase diagram, already discussed in section \ref{subsec:biaxialbackground} (see Fig.\ref{fig:SVDphasemap}). Here, except at $\lambda=0$, the ground state is always $N_B$. For small $\lambda$, the $N_B$ phase occurs at very low temperatures. The stability of $N_B$ gradually increases relative to that of $N_U$ on increasing $\lambda$, up to the triple point $\lambda 	\approx 0.33$, where all phases coexist. The tricritical point at $\lambda \approx 0.3$ separates the first-order and continuous  sections of the $N_B-N_U$ transition line. For $\lambda \agt 0.33$, a single first order $N_B-I$ transition replaces the phase sequence $N_B-N_U-I$. 

For $\alpha = 0.3$ (see Fig.\ref{fig:KLZSVDphasemapdelta00alpha03}) contains several features reminiscent of the GMA case.  Indeed, the topological structure of this phase diagram is very similar to its GMA countrepart in Fig.\ref{fig:KLZGMphasemapdelta00alpha03}. The main distinction is that here there is a line of first order $N_B-I$ transitions, which has shrunk in the GMA case to a single point.   

Specifically, the high temperature part of this diagram ($T^* \agt$ 0.2) is unchanged from the $\alpha = 0$ case. The orientational-translational coupling is still  too weak  to turn on the smectic modulation. Likewise, at sufficiently low temperatures, both uniaxial $\Sm A_U$ and biaxial smectic $\Sm A_B$ phases are stabilized, with the stability of the $\Sm A_B$ increasing gradually at larger $\lambda$.  The ground state $N_B$  is again replaced by  $\Sm A_B$ , while for values of $\lambda$ for which the $N_B$ phase requires low temperatures, the $N_U$ phase again yields to a $\Sm A_U$ phase. This picture again seems generic for low positive $\alpha$. The $N_B$ phase has been restricted to a smaller range of $0.13 \alt \lambda \alt 0.5$. The temperature at the $N_U-\Sm A_U$ phase boundary remains independent of $\alpha$, and the transition is continuous. Likewise, the $N_U-N_B$ transition remains continuous for $0.13 \alt \lambda \alt 0.3$, as are the new $N_B-\Sm A_B$ and $\Sm A_U- \Sm A_B$.  Finally, as in the GMA case, there is a four-phase coexistence point at $\lambda \approx 0.13, T^* \approx 0.12$, at which the three continuous phase transition lines $N_U-N_B$, $N_U-\Sm A_U$, $\Sm A_U-\Sm A_B$ collide.

For $\alpha = 0.9$ (see Fig.\ref{fig:KLZSVDphasemapdelta00alpha09}), whereas in the GMA case, there stilll remained a thin region of $N_B$  stability,  here the region of $N_B$ has been entirely overtaken by $\Sm A_B$. The stability region of $N_U$ is reduced into a narrow stripe at $T^* \approx 0.2$. The $\Sm A_U - N_U$ transition is now first order, consistent with the $\lambda = 0$ phase diagram in Fig.\ref{fig:KLZunidelta00}. As in the $\alpha = 0.3$ case, the $\Sm A_B$ phase is increasingly stable at higher $\lambda$. A tricritical point at $\lambda \approx 0.16$ separates  the first-order and continuous regions of the  $\Sm A_B-\Sm A_U$ boundary line, although the first order section of this line is very short.  
The point of contact  $\Sm A_B-N_U$ at $\alpha = 0.3$  is now transformed into a first order transition line for $0.17 \alt \lambda \alt 0.23$, followed by the new first-order $\Sm A_B-I$ transition line for $0.23 \alt \lambda \alt 0.5$.

In the final case we consider, $\alpha = 1.2$ (see Fig.\ref{fig:KLZSVDphasemapdelta00alpha12}), the entire nematic region has now been replaced by smectic phases. The $N_U-I$ transition line is replaced by a line of direct first-order $\Sm A_U-I$ transitions for $0 \leq \lambda \alt 0.15$. This feature of the phase diagram is consistent with the smectic molecular field theory  \cite{KLZ1985}, shown in Fig. \ref{fig:KLZunidelta00} for $\lambda = 0$. This $\Sm A_U-I$ transition is followed by a continuous $\Sm A_B- \Sm A_U$ transition at a lower temperature. This phase sequence is also identical to that in the GMA case for $\alpha=1.5$ (Fig.\ref{fig:KLZGMphasemapdelta00alpha15}) 

\begin{figure}[htb]
    \begin{center}
           \includegraphics[width=3.5in]{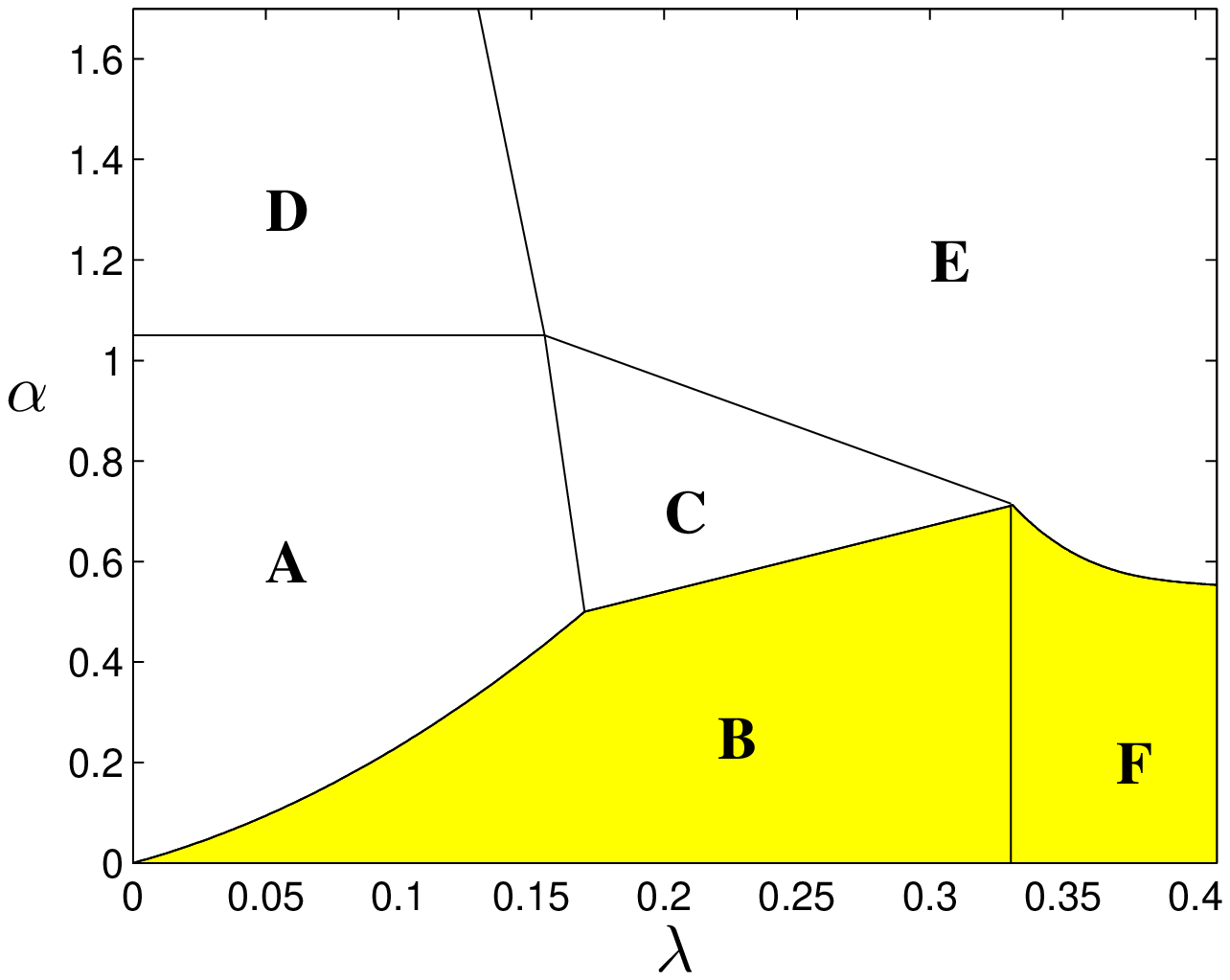}
    \end{center}
    \caption{SVD: phase sequences observed (approximate) in the $(\lambda,\alpha)$ (biaxiality-smecticity) plane. The regions with stabilized $N_B$ are shaded (yellow online). \newline 
		\begin{tabular}{llllllll}
		\textbf{A}:& $\Sm A_B-\Sm A_U-N_U-I~~~$&&  \textbf{B}:&$\Sm A_B-N_B-N_U-I$ \\
		\textbf{C}:& $\Sm A_B-N_U-I~~~$ &&  \textbf{D}:& $\Sm A_B-\Sm A_U-I$ \\
		\textbf{E}:&$\Sm A_B-I~~~$ &&  \textbf{F}:& $\Sm A_B-N_B-I$
		\end{tabular}
		}
\label{fig:SVDBiaxVSsmec}
\end{figure}

Fig.\ref{fig:SVDBiaxVSsmec} summarizes these results, showing  the phase sequences as a function of the two control parameters  $\lambda, \alpha$. There  are now six predicted phase sequences. Five are as in the GMA case (shaded regions $A$ through $E$).  Only  two shaded regions, \textbf{B} and the new \textbf{F}),  permit a biaxial nematic $N_B$, and these regions are restricted to a narrow region of the control parameter plane.  

\section{Discussion}
\label{sec:discussionsection}

We have developed a family of simplified two-parameter molecular field theories, which allow for both smectic and biaxial nematic phases.  The theories  extend  and combine standard models involving, on the one hand uniaxial nematic and smectic phases, and on the other hand  uniaxial and biaxial nematics.  We first discuss the approximations we have made.

The phase diagrams corresponding to the nematic limits of these models are slightly different. The geometric mean approximation (GMA) uses a Lorentz-Berthelot combination rule to reduce the number of molecular biaxiality parameters, and seems to correspond to model interaction potentials between some simple model slab-like molecules. The Sonnet-Virga-Durand parameterization is easier to use mathematically, but harder to justify physically, except in the case of molecules of complicated shape. In the former case, in the pure nematic limit, there is a single Landau multicritical point at which the uniaxial and biaxial nematic states touch the isotropic phase. In the latter case in the nematic limit there is a line of first order isotropic-biaxial nematic transitions. Luckily, however, more general theoretical studies \cite{Matteis2005b,turzi2013} suggest strongly that one-parameter slices through  general nematic biaxiality phase diagrams always give rise to phase portraits which possess the same topology as  one of our approximations (i.e. one of the diagrams shown in  Fig. \ref{fig:GMAvsSVD}).

The uniaxial smectic limit of this model involves the KLZ decoupling approximation \cite{KLZ1985}, in which the molecular orientational distributtion function is independent of position. Although this approximation cannot be completely accurate, it has great mathematical simplifying power. The main topological features of the phase diagram -- in particular the passage from a continuous to a first-order nematic-smectic transition with increasing translational-orientational coupling $\alpha$, followed by the elimination of the nematic phase at still higher $\alpha$ -- are correctly captured. In some cases  \cite{bates99,pizzirusso11}, simulation of  a model system has shown good agreement with this ansatz, while recent atomic simulations of 8CB, on the other hand, show that there are circumstances when this ansatz fails \cite{palermo2013}.     

In this paper we seek general explanations, rather than  close numerical agreement  with phase diagrams and thermodynamics  in particular cases.  The starting points for the models we have used in this paper are extremely simple. For our purposes they seem sufficiently  well-founded.   

The topologies of the phase diagrams predicted by our two zero-smecticity nematic approximation do differ. But nevertheless, subject to this proviso,  the resulting  emergent smectic biaxial phase diagrams do carry many features in common. For low values of the smectic parameter $\alpha$, the uniaxial and biaxial nematic phases  retain their integrity, but lower temperature phases are uniaxial and biaxial smectics; the ground state is always biaxial smectic. For higher values $\alpha$, the nematic phases, and in particular the biaxial nematic phase, are restricted to very narrow regions of the phase diagram, and eventually squeezed out altogether. 

A number of phase progressions have been predicted. The low smecticity $I-N_U-N_B$ (with decreasing temperature)  gives way to $I-N_U-N_B-\Sm A_B$ or $I-N_U-\Sm A_U-\Sm A_B$. The progression from the uniaxial nematic to the biaxial smectic seems to pass through either the biaxial nematic or the uniaxial smectic, but never both.  The SVD approximation permits a direct first-order $I-N_B$ at low $\alpha$, and hence a direct $I-N_B-\Sm A_B$ transition at higher $\alpha$.  But  of the predicted phase progressions, only the $I-N_B-\Sm A_B$ progression  does not occur in the simple geometric mean approximation.  As $\alpha$ is further increased, the nematic phases progressively disappear. In this regime, for higher biaxiality we predict a direct $I-\Sm A_B$ transition, while for lower biaxiality  we expect the more  indirect $I-\Sm A_U-\Sm A_B$ progression.   

The set of phase progressions have been summarized in Figs. \ref{fig:GMABiaxVSsmec} and  \ref{fig:SVDBiaxVSsmec} for the GMA and the SVD respectively. Here the six possible phase orderings found in our theory are shown in the biaxiality-smecticity plane.  The topological features of these diagrams are very similar.  High smecticity and low biaxiality together  predict the $I-\Sm A_U-\Sm A_B$ progression (regions $D$ in these figures). A lower degree of smecticity (regions $A$) allows the uniaxial nematic to intrude:  $I-N_U-\Sm A_U-\Sm A_B$ .  High biaxiality and high smecticity (regions $E$) predict a direct $I-\Sm A_B$ transition; now only the isotropic and biaxial smectic phases are allowed and all others have been squeezed out. For intermediate values of both parameters  (regions $C$), our results predict that there will be a region in which   there is a direct transition from the uniaxial nematic to the biaxial smectic ($I-N_U-\Sm A_B$. Both approximation schemes allow for a low biaxiality, low smecticity regime (region $B$) in which a biaxial nematic intrudes between the uniaxial nematic and the biaxial smectic ($I-N_U-N_B-\Sm A_B$).  The SVD approximation alone predicts a further region $F$, for low smecticity and high biaxiality, in which  the biaxial nematic occurs between the biaxial smectic and the isotropic phases: $I-N_B-\Sm A_B$. 

Of the six possible phase progressions, only two (regions B, common to both approximations,  and region  $F$ occurring only in the SVD approximation) include the $N_B$ phase. In each case these regions cover a relatively small proportion of the phase space. In addition, in the SVD case, where region $B$ extends into a region of low biaxiality, the biaxial nematic phase occurs at low temperatures and would surely be pre-empted by a crystalline phase.  The hypothesis that biaxial nematic phases are at least to some extent preempted by biaxial smectics thus seems to receive considerable informal support from  Figs. \ref{fig:GMABiaxVSsmec} and  \ref{fig:SVDBiaxVSsmec}.

\begin{figure*}[hbt]
    \begin{center}
        \subfloat[Dependence of scaled transition temperature on  smectic interaction parameter $\alpha$  (GMA: $\gamma = 0.24$)]{
           \includegraphics[width=0.35\textwidth]{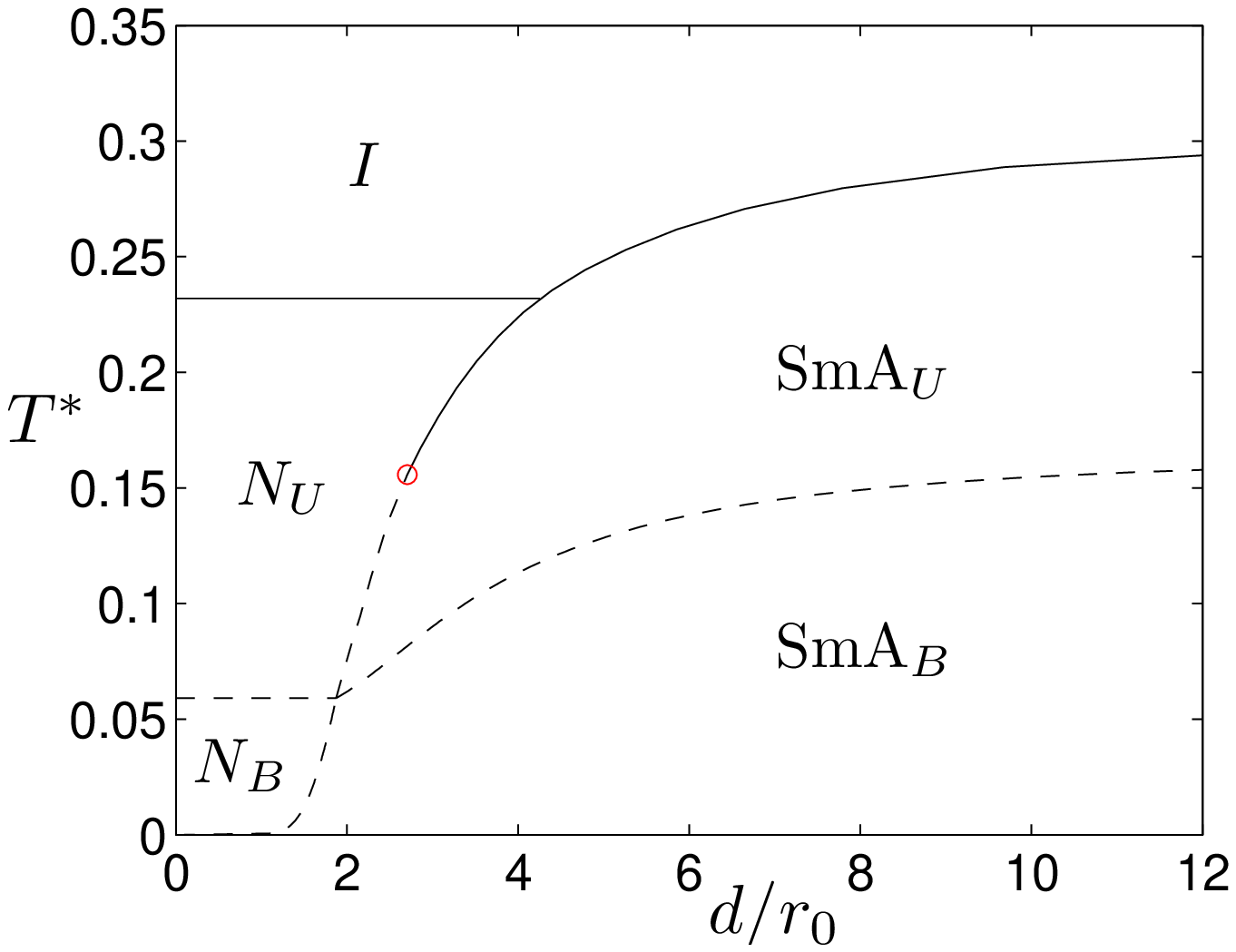}
           \label{fig:KLZGMphasemapdelta00gamma024}
        }
        \subfloat[Experimental results \cite{Sadashiva2004}. ]{%
            \includegraphics[width=0.35\textwidth]{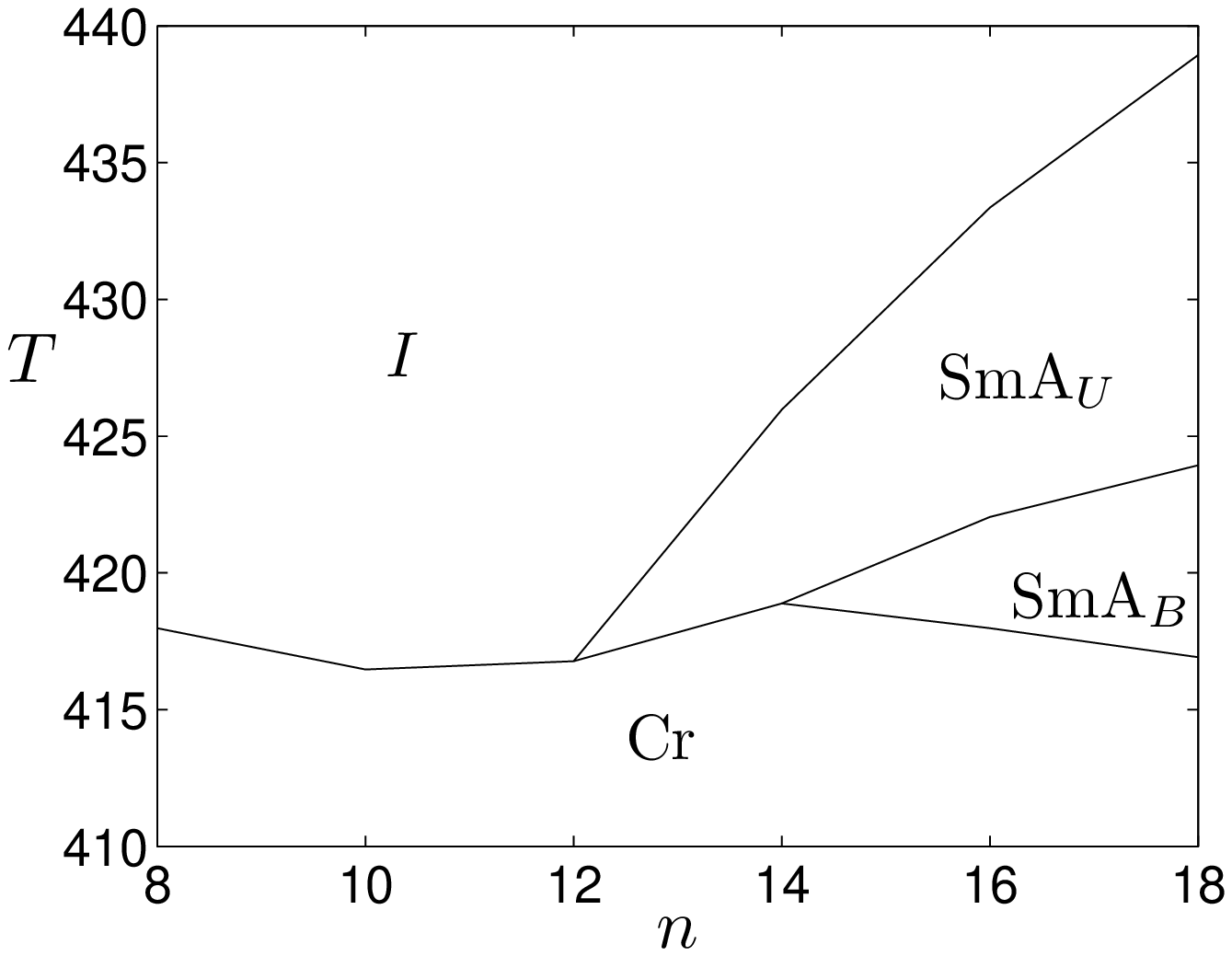}
            \label{fig:phasemapsadashiva2003}
        }
    \end{center}
    \caption{(b):  Dependence of  transition temperature on number of carbon atoms in the flexible chain of a bent-core molecule, after ref. [\onlinecite{Sadashiva2004}]. Antiferroelectric $\mathrm{SmA}_d$ is uniaxial;  $\mathrm{SmA}_dP_A$ is biaxial, $Cr$: crystal.
     }
    \label{fig:KLZBiaxSmecAphasemapsdelta00fixalpha}
\end{figure*}

Some more specific contact with experiment can be made  as follows. Approximate  values for parameters $\alpha$  and $\gamma$  can be derived  for  a series of rigid bent-core  molecule biaxial smectic A phases \cite{Sadashiva2002, Sadashiva2004},. For $\alpha$, we use  an argument due to McMillan \cite{McMillan1971}, and for $\gamma$, we use an argument due to Ferrarini \etal \cite{Ferrarini1996,Luckhurst2001}.

McMillan \cite{McMillan1971} proposed the following relation linking $\alpha$,  the length $r_0 $ of  a mesogenic unit in a smectogenic molecule, and the smectic layer spacing $d$: 
\begin{equation}
\label{eq:alphadef}
\alpha = 2 \exp{-\left( \frac{\pi r_0}{d} \right)^2},
\end{equation}
Eq.\eqref{eq:alphadef} can now be inverted, yielding 
\begin{equation}
\label{eq:dr0}
\frac{d}{r_0} = \frac{\pi}{\sqrt{\ln{2} - \ln{\alpha}}}.
\end{equation}
This gives the dependence of period spacing $d$ on $\alpha$. We now further suppose the smectic spacing $d$ to be  proportional to flexible chain length. It is now possible to  make contact with experiments, in which phase maps are constructed in a $T-n$ plane, where $n \sim d$ is flexible chain length. 

In the GMA,  the biaxiality parameter $\gamma$ has been related to the  interarm angle of bent-core molecules\cite{Ferrarini1996,Luckhurst2001} by the following equation: 
\begin{equation}
\label{BiaxialityVshaped}
\gamma = \sqrt{\frac{3}{2}}\left(\frac{1+\cos\theta}{1-3\cos{\theta}}\right).
\end{equation}
The chemical structure of the compound used in the experiments by Sadashiva \etal \cite{Sadashiva2004} suggests an interarm angle of 120$^0$, implying $\gamma \approx 0.24$. Thus, within the GMA, we can calculate a phase map for $\gamma = 0.24$ as a function of  $T^*$ and $d/r_0$. We show this phase diagram in Fig.\ref{fig:KLZBiaxSmecAphasemapsdelta00fixalpha}, together with the experimental results by Sadashiva \etal \cite{Sadashiva2004}.

The details of the theoretical and observed phase maps differ. We note in particular that  although the $N_U$ phase is absent in the experiments, it appears in the phase maps. On the other hand, however, some features of the two phase maps resemble each other.  Specifically, both the biaxial smectic A-to-uniaxial smectic A and uniaxial smectic A-to-isotropic phase transition temperatures increase with the number of carbon atoms in the flexible chain. Likewise, both these phase transition temperatures increase with the parameter $\alpha$ in the molecular field theory. 

On the other hand, in the experiments, the onset of $\Sm A_B$ occurs at much higher temperatures than the naive theory predicts. This may be because in the experiment, the smectic biaxial phase possesses  a bilayer layer structure, and also possesses antiferroelectric order. In addition, the molecules have dipolar shape. Thus transverse dipolar interaction may have stabilized the biaxial smectic phase. Future work,  explicitly including antiferroelectric order, may resolve this point. 

The smectic parameter $\alpha$ for uniaxial systems can be estimated by comparing  results for $T_{\Sm A_UN_U}/T_{N_UI}$ with experimental data of two calamitic series $n$CB \cite{thoen1984} and $\bar{n}$S5 \cite{marynissen1985}, where $n$ is the number of Carbon atoms in the flexible chain. The phase sequence $\Sm A_U-N_U-I$ occurs for $n=8,9$ for $n$CB  and $n=8,9,10$ for $\bar{n}$S5. The estimated value of $\alpha$ for 8CB and $\bar{8}$S5 are 1.02 and 0.92, respectively. These calamitic molecules can be interpreted in the GMA model as having $\gamma$ close to zero. From Fig.\ref{fig:GMABiaxVSsmec} we see that the molecules 8CB and $\bar{8}$S5 are in region {\bf A} of the parameter space.

We can also estimate $\alpha$ for shorter chain length. Using Eq.\eqref{eq:dr0} and supposing that $d/r_0 = Cn$, we found that $C$ decreases with $n$ for both $n$CB and $\bar{n}$S5 series; and this value for $\bar{n}$S5 is slightly smaller than $n$CB with the same $n$. Thus, we use $C=0.45$ (for $\bar{8}$S5) as the lower bound for smaller $n$ for a general calamitic molecule. Thus for the shortest chain length  for which a nematic phase exists $n=5$, the estimated value for $\alpha$, given by Eq.\eqref{eq:alphadef} is 0.28, which is clearly still in region {\bf A} in Fig.\ref{fig:GMABiaxVSsmec}. Note that for this value of $\alpha$ to fall into the region with stabilized $N_B$, we need $0.3 \alt \gamma \leq 1/\sqrt{6} $.  This is a rather high degree of biaxiality, and for higher $n$, the biaxial nematic region would be further restricted.  By way of example,  in the case of symmetric bent-core molecules with $n=5$, using eq.\eqref{BiaxialityVshaped},  an interarm angle of $109.47^0 \leq \gamma \alt 115^0$ would be required to stabilize the $N_B$ phase.

We also note that some features of  our results may not be generic, and may  not persist when the KLZ decoupling approximation is relaxed.  Some examples are: 
\begin{itemize}
\item[(a)] Both Figs.\ref{fig:KLZGMphasemapsdelta00} and \ref{fig:KLZSVDphasemapsdelta00} include what seem to be exceptional points of four-phase contact ($N_U-N_B-\Sm A_U-\Sm A_B$., where the continuous  $N_U-N_B$  and $N_U-\Sm A_U$ lines collide. The fate of these points when the molecular field approximations include better chemical detail, or if fluctuations are included, is not clear. 
\item[(b)] In the present theory, the $N_U-\Sm A_U$ tricritical lines appear in phase map slices in the smecticity-temperature plane,  but do not appear in the Figs.\ref{fig:KLZGMphasemapsdelta00} and \ref{fig:KLZSVDphasemapsdelta00}, which present phase map slices in the biaxiality-temperature plane.  Presumably in more accurate theories, the tricritical points would be visible in both sets of slices. 
\item[(c)] In the phase progression maps, Figs \ref{fig:GMABiaxVSsmec} and  \ref{fig:SVDBiaxVSsmec},  both include points where four sets of phase progressions touch. One would not normally expect three lines in a picture to collide at a single point, although there may be thermodynamic reasons why this cannot be avoided. 
\end{itemize} 
 
In summary, we have developed a molecular field theory for biaxial smectic A phase by combining McMillan theory for uniaxial smectic A phases \cite{McMillan1971} and  a generalized version of Straley theory \cite{Straley1974} for biaxial nematics. By applying several approximations, the number of input parameters in the model is restricted to a single degree of smecticity  and a single degree of molecular biaxiality.   Likewise, the number of order parameters is reduced from nine to three.  

Numerical results show that the stability of smectic A phases increase on increasing the smectic interaction $\alpha$. A system with high $\alpha$ and small biaxiality can still form a biaxial smectic A phase at high temperature. In contrast, for the same biaxiality, a system with low smectic interaction forms a biaxial nematic phase only at unphysically  low temperatures.  Our results also agree with empirical evidence that the stability of the biaxial and uniaxial smectic A phases increases with flexible chain length. On the basis of this study,  we conclude that, with the same molecular biaxiality, macroscopic biaxial ordering is easier to form in the smectic A phases than in the nematic phases.

\begin{acknowledgments}
TBTT  acknowledges financial support of a School Ph.D. Studentship  from the School of Mathematical Sciences, University of Southampton. We thank E.G. Virga, O.D. Lavrentovich,  M.A. Osipov, I.I. Smalyukh  and P.I.C. Teixeira  for helpful discussions. TJS and TBTT thank the Isaac Newton Institute for Mathematical Sciences,  University of Cambridge, for hospitality while some of the final parts of this reseach were carried out.  
\end{acknowledgments}

\bibliography{smb}

\begin{thebibliography}{44}%
\makeatletter
\providecommand \@ifxundefined [1]{%
 \@ifx{#1\undefined}
}%
\providecommand \@ifnum [1]{%
 \ifnum #1\expandafter \@firstoftwo
 \else \expandafter \@secondoftwo
 \fi
}%
\providecommand \@ifx [1]{%
 \ifx #1\expandafter \@firstoftwo
 \else \expandafter \@secondoftwo
 \fi
}%
\providecommand \natexlab [1]{#1}%
\providecommand \enquote  [1]{``#1''}%
\providecommand \bibnamefont  [1]{#1}%
\providecommand \bibfnamefont [1]{#1}%
\providecommand \citenamefont [1]{#1}%
\providecommand \href@noop [0]{\@secondoftwo}%
\providecommand \href [0]{\begingroup \@sanitize@url \@href}%
\providecommand \@href[1]{\@@startlink{#1}\@@href}%
\providecommand \@@href[1]{\endgroup#1\@@endlink}%
\providecommand \@sanitize@url [0]{\catcode `\\12\catcode `\$12\catcode
  `\&12\catcode `\#12\catcode `\^12\catcode `\_12\catcode `\%12\relax}%
\providecommand \@@startlink[1]{}%
\providecommand \@@endlink[0]{}%
\providecommand \url  [0]{\begingroup\@sanitize@url \@url }%
\providecommand \@url [1]{\endgroup\@href {#1}{\urlprefix }}%
\providecommand \urlprefix  [0]{URL }%
\providecommand \Eprint [0]{\href }%
\providecommand \doibase [0]{http://dx.doi.org/}%
\providecommand \selectlanguage [0]{\@gobble}%
\providecommand \bibinfo  [0]{\@secondoftwo}%
\providecommand \bibfield  [0]{\@secondoftwo}%
\providecommand \translation [1]{[#1]}%
\providecommand \BibitemOpen [0]{}%
\providecommand \bibitemStop [0]{}%
\providecommand \bibitemNoStop [0]{.\EOS\space}%
\providecommand \EOS [0]{\spacefactor3000\relax}%
\providecommand \BibitemShut  [1]{\csname bibitem#1\endcsname}%
\let\auto@bib@innerbib\@empty
\bibitem [{\citenamefont {Freiser}(1970)}]{freiser70}%
  \BibitemOpen
  \bibfield  {author} {\bibinfo {author} {\bibfnamefont {M.~J.}\ \bibnamefont
  {Freiser}},\ }\href@noop {} {\bibfield  {journal} {\bibinfo  {journal} {Phys.
  Rev. Lett.}\ }\textbf {\bibinfo {volume} {24}},\ \bibinfo {pages} {1041}
  (\bibinfo {year} {1970})}\BibitemShut {NoStop}%
\bibitem [{\citenamefont {Boccara}, \citenamefont {Mejdani},\ and\
  \citenamefont {de~Seze}(1977)}]{Boccara1976}%
  \BibitemOpen
  \bibfield  {author} {\bibinfo {author} {\bibfnamefont {N.}~\bibnamefont
  {Boccara}}, \bibinfo {author} {\bibfnamefont {R.}~\bibnamefont {Mejdani}}, \
  and\ \bibinfo {author} {\bibfnamefont {L.}~\bibnamefont {de~Seze}},\
  }\href@noop {} {\bibfield  {journal} {\bibinfo  {journal} {J. de Phys.}\
  }\textbf {\bibinfo {volume} {7}},\ \bibinfo {pages} {149} (\bibinfo {year}
  {1977})}\BibitemShut {NoStop}%
\bibitem [{\citenamefont {Remler}\ and\ \citenamefont
  {Haymet}(1986)}]{RemlerHaymet1986}%
  \BibitemOpen
  \bibfield  {author} {\bibinfo {author} {\bibfnamefont {D.~K.}\ \bibnamefont
  {Remler}}\ and\ \bibinfo {author} {\bibfnamefont {A.~D.~J.}\ \bibnamefont
  {Haymet}},\ }\href@noop {} {\bibfield  {journal} {\bibinfo  {journal} {J.
  Phys. Chem.}\ }\textbf {\bibinfo {volume} {90}},\ \bibinfo {pages} {5426}
  (\bibinfo {year} {1986})}\BibitemShut {NoStop}%
\bibitem [{\citenamefont {Rosso}(2007)}]{Rosso2007}%
  \BibitemOpen
  \bibfield  {author} {\bibinfo {author} {\bibfnamefont {R.}~\bibnamefont
  {Rosso}},\ }\href@noop {} {\bibfield  {journal} {\bibinfo  {journal} {Liquid
  Crystals}\ }\textbf {\bibinfo {volume} {34}},\ \bibinfo {pages} {737}
  (\bibinfo {year} {2007})}\BibitemShut {NoStop}%
\bibitem [{\citenamefont {Berardi}\ \emph {et~al.}(2008)\citenamefont
  {Berardi}, \citenamefont {Muccioli}, \citenamefont {Orlandi}, \citenamefont
  {Ricci},\ and\ \citenamefont {Zannoni}}]{berardi2008}%
  \BibitemOpen
  \bibfield  {author} {\bibinfo {author} {\bibfnamefont {R.}~\bibnamefont
  {Berardi}}, \bibinfo {author} {\bibfnamefont {L.}~\bibnamefont {Muccioli}},
  \bibinfo {author} {\bibfnamefont {S.}~\bibnamefont {Orlandi}}, \bibinfo
  {author} {\bibfnamefont {M.}~\bibnamefont {Ricci}}, \ and\ \bibinfo {author}
  {\bibfnamefont {C.}~\bibnamefont {Zannoni}},\ }\href@noop {} {\bibfield
  {journal} {\bibinfo  {journal} {J. Phys.: Condensed Matter}\ }\textbf
  {\bibinfo {volume} {20}},\ \bibinfo {pages} {463101} (\bibinfo {year}
  {2008})}\BibitemShut {NoStop}%
\bibitem [{\citenamefont {Luckhurst}(2004)}]{Luckhurst2004}%
  \BibitemOpen
  \bibfield  {author} {\bibinfo {author} {\bibfnamefont {G.~R.}\ \bibnamefont
  {Luckhurst}},\ }\href@noop {} {\bibfield  {journal} {\bibinfo  {journal}
  {Nature}\ }\textbf {\bibinfo {volume} {430}},\ \bibinfo {pages} {413}
  (\bibinfo {year} {2004})}\BibitemShut {NoStop}%
\bibitem [{\citenamefont {Madsen}\ \emph {et~al.}(2004)\citenamefont {Madsen},
  \citenamefont {Dingemans}, \citenamefont {Nakata},\ and\ \citenamefont
  {Samulski}}]{Madsen2004}%
  \BibitemOpen
  \bibfield  {author} {\bibinfo {author} {\bibfnamefont {L.~A.}\ \bibnamefont
  {Madsen}}, \bibinfo {author} {\bibfnamefont {T.~J.}\ \bibnamefont
  {Dingemans}}, \bibinfo {author} {\bibfnamefont {M.}~\bibnamefont {Nakata}}, \
  and\ \bibinfo {author} {\bibfnamefont {E.~T.}\ \bibnamefont {Samulski}},\
  }\href@noop {} {\bibfield  {journal} {\bibinfo  {journal} {Phys. Rev. Lett.}\
  }\textbf {\bibinfo {volume} {92}},\ \bibinfo {pages} {145505} (\bibinfo
  {year} {2004})}\BibitemShut {NoStop}%
\bibitem [{\citenamefont {Acharya}, \citenamefont {Primak},\ and\ \citenamefont
  {Kumar}(2004)}]{Kumar2004}%
  \BibitemOpen
  \bibfield  {author} {\bibinfo {author} {\bibfnamefont {B.~R.}\ \bibnamefont
  {Acharya}}, \bibinfo {author} {\bibfnamefont {A.}~\bibnamefont {Primak}}, \
  and\ \bibinfo {author} {\bibfnamefont {S.}~\bibnamefont {Kumar}},\
  }\href@noop {} {\bibfield  {journal} {\bibinfo  {journal} {Phys. Rev. Lett.}\
  }\textbf {\bibinfo {volume} {92}},\ \bibinfo {pages} {145506} (\bibinfo
  {year} {2004})}\BibitemShut {NoStop}%
\bibitem [{\citenamefont {Yu}\ and\ \citenamefont {Saupe}(1980)}]{YuSaupe1980}%
  \BibitemOpen
  \bibfield  {author} {\bibinfo {author} {\bibfnamefont {L.~J.}\ \bibnamefont
  {Yu}}\ and\ \bibinfo {author} {\bibfnamefont {A.}~\bibnamefont {Saupe}},\
  }\href@noop {} {\bibfield  {journal} {\bibinfo  {journal} {Phys. Rev. Lett.}\
  }\textbf {\bibinfo {volume} {45}},\ \bibinfo {pages} {1000} (\bibinfo {year}
  {1980})}\BibitemShut {NoStop}%
\bibitem [{\citenamefont {G\"{o}rtz}\ and\ \citenamefont
  {Goodby}(2005)}]{GoertzGoodby2005}%
  \BibitemOpen
  \bibfield  {author} {\bibinfo {author} {\bibfnamefont {V.}~\bibnamefont
  {G\"{o}rtz}}\ and\ \bibinfo {author} {\bibfnamefont {J.~W.}\ \bibnamefont
  {Goodby}},\ }\href@noop {} {\bibfield  {journal} {\bibinfo  {journal} {Chem.
  Commun.}\ ,\ \bibinfo {pages} {3262}} (\bibinfo {year} {2005})}\BibitemShut
  {NoStop}%
\bibitem [{\citenamefont {Senyuk}\ \emph {et~al.}(2010)\citenamefont {Senyuk},
  \citenamefont {Wonderly}, \citenamefont {Mathews}, \citenamefont {Li},
  \citenamefont {Shiyanovskii},\ and\ \citenamefont
  {Lavrentovich}}]{Senyuk2010}%
  \BibitemOpen
  \bibfield  {author} {\bibinfo {author} {\bibfnamefont {B.}~\bibnamefont
  {Senyuk}}, \bibinfo {author} {\bibfnamefont {H.}~\bibnamefont {Wonderly}},
  \bibinfo {author} {\bibfnamefont {M.}~\bibnamefont {Mathews}}, \bibinfo
  {author} {\bibfnamefont {Q.}~\bibnamefont {Li}}, \bibinfo {author}
  {\bibfnamefont {S.~V.}\ \bibnamefont {Shiyanovskii}}, \ and\ \bibinfo
  {author} {\bibfnamefont {O.~D.}\ \bibnamefont {Lavrentovich}},\ }\href@noop
  {} {\bibfield  {journal} {\bibinfo  {journal} {Phys. Rev. E}\ }\textbf
  {\bibinfo {volume} {82}},\ \bibinfo {pages} {041711} (\bibinfo {year}
  {2010})}\BibitemShut {NoStop}%
\bibitem [{\citenamefont {Kim}\ \emph {et~al.}(2013)\citenamefont {Kim},
  \citenamefont {Majumdar}, \citenamefont {Senyuk}, \citenamefont {Tortora},
  \citenamefont {Seltmann}, \citenamefont {Lehmann}, \citenamefont {J\'akli},
  \citenamefont {Gleeeson}, \citenamefont {Lavrentovich},\ and\ \citenamefont
  {Sprunt}}]{Kim2013}%
  \BibitemOpen
  \bibfield  {author} {\bibinfo {author} {\bibfnamefont {Y.~K.}\ \bibnamefont
  {Kim}}, \bibinfo {author} {\bibfnamefont {M.}~\bibnamefont {Majumdar}},
  \bibinfo {author} {\bibfnamefont {B.~I.}\ \bibnamefont {Senyuk}}, \bibinfo
  {author} {\bibfnamefont {L.}~\bibnamefont {Tortora}}, \bibinfo {author}
  {\bibfnamefont {J.}~\bibnamefont {Seltmann}}, \bibinfo {author}
  {\bibfnamefont {M.}~\bibnamefont {Lehmann}}, \bibinfo {author} {\bibfnamefont
  {A.}~\bibnamefont {J\'akli}}, \bibinfo {author} {\bibfnamefont
  {J.}~\bibnamefont {Gleeeson}}, \bibinfo {author} {\bibfnamefont {O.~D.}\
  \bibnamefont {Lavrentovich}}, \ and\ \bibinfo {author} {\bibfnamefont
  {S.}~\bibnamefont {Sprunt}},\ }\href@noop {} {\bibfield  {journal} {\bibinfo
  {journal} {Soft Matter}\ }\textbf {\bibinfo {volume} {8}},\ \bibinfo {pages}
  {8880} (\bibinfo {year} {2013})}\BibitemShut {NoStop}%
\bibitem [{\citenamefont {Galindo}\ \emph {et~al.}(2003)\citenamefont
  {Galindo}, \citenamefont {Haslam}, \citenamefont {Varga}, \citenamefont
  {Jackson}, \citenamefont {Vanakaras}, \citenamefont {Photinos},\ and\
  \citenamefont {Dunmur}}]{Galindo2003}%
  \BibitemOpen
  \bibfield  {author} {\bibinfo {author} {\bibfnamefont {A.}~\bibnamefont
  {Galindo}}, \bibinfo {author} {\bibfnamefont {A.~J.}\ \bibnamefont {Haslam}},
  \bibinfo {author} {\bibfnamefont {S.}~\bibnamefont {Varga}}, \bibinfo
  {author} {\bibfnamefont {G.}~\bibnamefont {Jackson}}, \bibinfo {author}
  {\bibfnamefont {A.}~\bibnamefont {Vanakaras}}, \bibinfo {author}
  {\bibfnamefont {D.~J.}\ \bibnamefont {Photinos}}, \ and\ \bibinfo {author}
  {\bibfnamefont {D.}~\bibnamefont {Dunmur}},\ }\href@noop {} {\bibfield
  {journal} {\bibinfo  {journal} {J. Chem. Phys.}\ }\textbf {\bibinfo {volume}
  {119}},\ \bibinfo {pages} {5216} (\bibinfo {year} {2003})}\BibitemShut
  {NoStop}%
\bibitem [{\citenamefont {Teixeira}, \citenamefont {Osipov},\ and\
  \citenamefont {Luckhurst}(2006)}]{Teixeira2006}%
  \BibitemOpen
  \bibfield  {author} {\bibinfo {author} {\bibfnamefont {P.~I.~C.}\
  \bibnamefont {Teixeira}}, \bibinfo {author} {\bibfnamefont {M.~A.}\
  \bibnamefont {Osipov}}, \ and\ \bibinfo {author} {\bibfnamefont {G.~R.}\
  \bibnamefont {Luckhurst}},\ }\href@noop {} {\bibfield  {journal} {\bibinfo
  {journal} {Phys. Rev. E}\ }\textbf {\bibinfo {volume} {73}},\ \bibinfo
  {pages} {061708} (\bibinfo {year} {2006})}\BibitemShut {NoStop}%
\bibitem [{\citenamefont {Matsushita}(1981)}]{Matsushita1981}%
  \BibitemOpen
  \bibfield  {author} {\bibinfo {author} {\bibfnamefont {M.}~\bibnamefont
  {Matsushita}},\ }\href@noop {} {\bibfield  {journal} {\bibinfo  {journal}
  {Mol. Cryst. Liq. Cryst.}\ }\textbf {\bibinfo {volume} {68}},\ \bibinfo
  {pages} {949} (\bibinfo {year} {1981})}\BibitemShut {NoStop}%
\bibitem [{\citenamefont {Hegmann}\ \emph {et~al.}(2001)\citenamefont
  {Hegmann}, \citenamefont {Kain}, \citenamefont {Diele}, \citenamefont
  {Pelzl},\ and\ \citenamefont {Tschierske}}]{Hegmann2001}%
  \BibitemOpen
  \bibfield  {author} {\bibinfo {author} {\bibfnamefont {T.}~\bibnamefont
  {Hegmann}}, \bibinfo {author} {\bibfnamefont {J.}~\bibnamefont {Kain}},
  \bibinfo {author} {\bibfnamefont {S.}~\bibnamefont {Diele}}, \bibinfo
  {author} {\bibfnamefont {G.}~\bibnamefont {Pelzl}}, \ and\ \bibinfo {author}
  {\bibfnamefont {C.}~\bibnamefont {Tschierske}},\ }\href@noop {} {\bibfield
  {journal} {\bibinfo  {journal} {Angew. Chem. Int. Ed.}\ }\textbf {\bibinfo
  {volume} {40}},\ \bibinfo {pages} {887} (\bibinfo {year} {2001})}\BibitemShut
  {NoStop}%
\bibitem [{\citenamefont {Kaznacheev}\ and\ \citenamefont
  {Hegmann}(2007)}]{Hegmann2007}%
  \BibitemOpen
  \bibfield  {author} {\bibinfo {author} {\bibfnamefont {K.}~\bibnamefont
  {Kaznacheev}}\ and\ \bibinfo {author} {\bibfnamefont {T.}~\bibnamefont
  {Hegmann}},\ }\href@noop {} {\bibfield  {journal} {\bibinfo  {journal} {Phys.
  Chem. Chem. Phys.}\ }\textbf {\bibinfo {volume} {9}},\ \bibinfo {pages}
  {1705} (\bibinfo {year} {2007})}\BibitemShut {NoStop}%
\bibitem [{\citenamefont {Yelamaggad}\ \emph {et~al.}(2004)\citenamefont
  {Yelamaggad}, \citenamefont {Prasad}, \citenamefont {Nair}, \citenamefont
  {Shashikala}, \citenamefont {Rao}, \citenamefont {Lobo},\ and\ \citenamefont
  {Chandrasekhar}}]{Yelamaggad2004}%
  \BibitemOpen
  \bibfield  {author} {\bibinfo {author} {\bibfnamefont {C.~V.}\ \bibnamefont
  {Yelamaggad}}, \bibinfo {author} {\bibfnamefont {S.~K.}\ \bibnamefont
  {Prasad}}, \bibinfo {author} {\bibfnamefont {G.~G.}\ \bibnamefont {Nair}},
  \bibinfo {author} {\bibfnamefont {I.~S.}\ \bibnamefont {Shashikala}},
  \bibinfo {author} {\bibfnamefont {D.~S.~S.}\ \bibnamefont {Rao}}, \bibinfo
  {author} {\bibfnamefont {C.~V.}\ \bibnamefont {Lobo}}, \ and\ \bibinfo
  {author} {\bibfnamefont {S.}~\bibnamefont {Chandrasekhar}},\ }\href@noop {}
  {\bibfield  {journal} {\bibinfo  {journal} {Angew. Chem. Int. Ed.}\ }\textbf
  {\bibinfo {volume} {43}},\ \bibinfo {pages} {3429} (\bibinfo {year}
  {2004})}\BibitemShut {NoStop}%
\bibitem [{\citenamefont {Wang}\ \emph {et~al.}(2012)\citenamefont {Wang},
  \citenamefont {Yoon}, \citenamefont {Bisoyi}, \citenamefont {Kumar},\ and\
  \citenamefont {Li}}]{Wang2012}%
  \BibitemOpen
  \bibfield  {author} {\bibinfo {author} {\bibfnamefont {Y.}~\bibnamefont
  {Wang}}, \bibinfo {author} {\bibfnamefont {H.~G.}\ \bibnamefont {Yoon}},
  \bibinfo {author} {\bibfnamefont {H.~K.}\ \bibnamefont {Bisoyi}}, \bibinfo
  {author} {\bibfnamefont {S.}~\bibnamefont {Kumar}}, \ and\ \bibinfo {author}
  {\bibfnamefont {Q.}~\bibnamefont {Li}},\ }\href@noop {} {\bibfield  {journal}
  {\bibinfo  {journal} {J. Mater. Chem.}\ }\textbf {\bibinfo {volume} {22}},\
  \bibinfo {pages} {20363} (\bibinfo {year} {2012})}\BibitemShut {NoStop}%
\bibitem [{\citenamefont {Eremin}\ \emph {et~al.}(2001)\citenamefont {Eremin},
  \citenamefont {Diele}, \citenamefont {Pelzl}, \citenamefont
  {N$\mathrm{\acute{a}}$dasi}, \citenamefont {Weissflog}, \citenamefont
  {Salfetnikova},\ and\ \citenamefont {Kresse}}]{Eremin2001}%
  \BibitemOpen
  \bibfield  {author} {\bibinfo {author} {\bibfnamefont {A.}~\bibnamefont
  {Eremin}}, \bibinfo {author} {\bibfnamefont {S.}~\bibnamefont {Diele}},
  \bibinfo {author} {\bibfnamefont {G.}~\bibnamefont {Pelzl}}, \bibinfo
  {author} {\bibfnamefont {H.}~\bibnamefont {N$\mathrm{\acute{a}}$dasi}},
  \bibinfo {author} {\bibfnamefont {W.}~\bibnamefont {Weissflog}}, \bibinfo
  {author} {\bibfnamefont {J.}~\bibnamefont {Salfetnikova}}, \ and\ \bibinfo
  {author} {\bibfnamefont {H.}~\bibnamefont {Kresse}},\ }\href@noop {}
  {\bibfield  {journal} {\bibinfo  {journal} {Phys. Rev. E}\ }\textbf {\bibinfo
  {volume} {64}},\ \bibinfo {pages} {051707} (\bibinfo {year}
  {2001})}\BibitemShut {NoStop}%
\bibitem [{\citenamefont {Reddy}\ \emph {et~al.}(2011)\citenamefont {Reddy},
  \citenamefont {Zhu}, \citenamefont {Shao}, \citenamefont {Korblova},
  \citenamefont {Gong}, \citenamefont {Shen}, \citenamefont {Garcia},
  \citenamefont {Glaser}, \citenamefont {Maclennan}, \citenamefont {Walba},\
  and\ \citenamefont {Clark}}]{ReddyScience2011}%
  \BibitemOpen
  \bibfield  {author} {\bibinfo {author} {\bibfnamefont {R.~A.}\ \bibnamefont
  {Reddy}}, \bibinfo {author} {\bibfnamefont {C.}~\bibnamefont {Zhu}}, \bibinfo
  {author} {\bibfnamefont {R.}~\bibnamefont {Shao}}, \bibinfo {author}
  {\bibfnamefont {E.}~\bibnamefont {Korblova}}, \bibinfo {author}
  {\bibfnamefont {T.}~\bibnamefont {Gong}}, \bibinfo {author} {\bibfnamefont
  {Y.}~\bibnamefont {Shen}}, \bibinfo {author} {\bibfnamefont {E.}~\bibnamefont
  {Garcia}}, \bibinfo {author} {\bibfnamefont {M.~A.}\ \bibnamefont {Glaser}},
  \bibinfo {author} {\bibfnamefont {J.~E.}\ \bibnamefont {Maclennan}}, \bibinfo
  {author} {\bibfnamefont {D.~M.}\ \bibnamefont {Walba}}, \ and\ \bibinfo
  {author} {\bibfnamefont {N.~A.}\ \bibnamefont {Clark}},\ }\href@noop {}
  {\bibfield  {journal} {\bibinfo  {journal} {Science}\ }\textbf {\bibinfo
  {volume} {332}},\ \bibinfo {pages} {72} (\bibinfo {year} {2011})}\BibitemShut
  {NoStop}%
\bibitem [{\citenamefont {Sadashiva}\ \emph {et~al.}(2002)\citenamefont
  {Sadashiva}, \citenamefont {Reddy}, \citenamefont {Pratibha},\ and\
  \citenamefont {Madhusudana}}]{Sadashiva2002}%
  \BibitemOpen
  \bibfield  {author} {\bibinfo {author} {\bibfnamefont {B.~K.}\ \bibnamefont
  {Sadashiva}}, \bibinfo {author} {\bibfnamefont {R.~A.}\ \bibnamefont
  {Reddy}}, \bibinfo {author} {\bibfnamefont {R.}~\bibnamefont {Pratibha}}, \
  and\ \bibinfo {author} {\bibfnamefont {N.~V.}\ \bibnamefont {Madhusudana}},\
  }\href@noop {} {\bibfield  {journal} {\bibinfo  {journal} {J. Mater. Chem.}\
  }\textbf {\bibinfo {volume} {12}},\ \bibinfo {pages} {943} (\bibinfo {year}
  {2002})}\BibitemShut {NoStop}%
\bibitem [{\citenamefont {Murthy}\ and\ \citenamefont
  {Sadashiva}(2004)}]{Sadashiva2004}%
  \BibitemOpen
  \bibfield  {author} {\bibinfo {author} {\bibfnamefont {H.~N.~S.}\
  \bibnamefont {Murthy}}\ and\ \bibinfo {author} {\bibfnamefont {B.~K.}\
  \bibnamefont {Sadashiva}},\ }\href@noop {} {\bibfield  {journal} {\bibinfo
  {journal} {Liq. Cryst.}\ }\textbf {\bibinfo {volume} {31}},\ \bibinfo {pages}
  {567} (\bibinfo {year} {2004})}\BibitemShut {NoStop}%
\bibitem [{\citenamefont {Yelamaggad}\ \emph {et~al.}(2006)\citenamefont
  {Yelamaggad}, \citenamefont {Shashikala}, \citenamefont {Rao}, \citenamefont
  {Nair},\ and\ \citenamefont {Prasad}}]{Yelamaggad2006}%
  \BibitemOpen
  \bibfield  {author} {\bibinfo {author} {\bibfnamefont {C.~V.}\ \bibnamefont
  {Yelamaggad}}, \bibinfo {author} {\bibfnamefont {I.~S.}\ \bibnamefont
  {Shashikala}}, \bibinfo {author} {\bibfnamefont {D.~S.~S.}\ \bibnamefont
  {Rao}}, \bibinfo {author} {\bibfnamefont {G.~G.}\ \bibnamefont {Nair}}, \
  and\ \bibinfo {author} {\bibfnamefont {S.~K.}\ \bibnamefont {Prasad}},\
  }\href@noop {} {\bibfield  {journal} {\bibinfo  {journal} {J. Mater. Chem.}\
  }\textbf {\bibinfo {volume} {16}},\ \bibinfo {pages} {4099} (\bibinfo {year}
  {2006})}\BibitemShut {NoStop}%
\bibitem [{\citenamefont {Berardi}\ and\ \citenamefont
  {Zannoni}(2000)}]{ZannoniBSmA2000}%
  \BibitemOpen
  \bibfield  {author} {\bibinfo {author} {\bibfnamefont {R.}~\bibnamefont
  {Berardi}}\ and\ \bibinfo {author} {\bibfnamefont {C.}~\bibnamefont
  {Zannoni}},\ }\href@noop {} {\bibfield  {journal} {\bibinfo  {journal} {J.
  Chem. Phys.}\ }\textbf {\bibinfo {volume} {113}},\ \bibinfo {pages} {5971}
  (\bibinfo {year} {2000})}\BibitemShut {NoStop}%
\bibitem [{\citenamefont {McMillan}(1971)}]{McMillan1971}%
  \BibitemOpen
  \bibfield  {author} {\bibinfo {author} {\bibfnamefont {W.~L.}\ \bibnamefont
  {McMillan}},\ }\href@noop {} {\bibfield  {journal} {\bibinfo  {journal}
  {Phys. Rev. A}\ }\textbf {\bibinfo {volume} {4}},\ \bibinfo {pages} {1238}
  (\bibinfo {year} {1971})}\BibitemShut {NoStop}%
\bibitem [{\citenamefont {McMillan}(1972)}]{McMillan1972}%
  \BibitemOpen
  \bibfield  {author} {\bibinfo {author} {\bibfnamefont {W.~L.}\ \bibnamefont
  {McMillan}},\ }\href@noop {} {\bibfield  {journal} {\bibinfo  {journal}
  {Phys. Rev. A}\ }\textbf {\bibinfo {volume} {6}},\ \bibinfo {pages} {936}
  (\bibinfo {year} {1972})}\BibitemShut {NoStop}%
\bibitem [{\citenamefont {Kventsel}, \citenamefont {Luckhurst},\ and\
  \citenamefont {Zewdie}(1985)}]{KLZ1985}%
  \BibitemOpen
  \bibfield  {author} {\bibinfo {author} {\bibfnamefont {G.~F.}\ \bibnamefont
  {Kventsel}}, \bibinfo {author} {\bibfnamefont {G.~R.}\ \bibnamefont
  {Luckhurst}}, \ and\ \bibinfo {author} {\bibfnamefont {H.~B.}\ \bibnamefont
  {Zewdie}},\ }\href@noop {} {\bibfield  {journal} {\bibinfo  {journal} {Mol.
  Phys.}\ }\textbf {\bibinfo {volume} {56}},\ \bibinfo {pages} {589} (\bibinfo
  {year} {1985})}\BibitemShut {NoStop}%
\bibitem [{\citenamefont {Luckhurst}\ \emph {et~al.}(1975)\citenamefont
  {Luckhurst}, \citenamefont {Zannoni}, \citenamefont {Nordio},\ and\
  \citenamefont {Segre}}]{Luckhurst1975}%
  \BibitemOpen
  \bibfield  {author} {\bibinfo {author} {\bibfnamefont {G.~R.}\ \bibnamefont
  {Luckhurst}}, \bibinfo {author} {\bibfnamefont {C.}~\bibnamefont {Zannoni}},
  \bibinfo {author} {\bibfnamefont {P.~L.}\ \bibnamefont {Nordio}}, \ and\
  \bibinfo {author} {\bibfnamefont {U.}~\bibnamefont {Segre}},\ }\href@noop {}
  {\bibfield  {journal} {\bibinfo  {journal} {Mol. Phys.}\ }\textbf {\bibinfo
  {volume} {30}},\ \bibinfo {pages} {1345} (\bibinfo {year}
  {1975})}\BibitemShut {NoStop}%
\bibitem [{\citenamefont {Sonnet}, \citenamefont {Virga},\ and\ \citenamefont
  {Durand}(2003)}]{Virga2003}%
  \BibitemOpen
  \bibfield  {author} {\bibinfo {author} {\bibfnamefont {A.}~\bibnamefont
  {Sonnet}}, \bibinfo {author} {\bibfnamefont {E.~G.}\ \bibnamefont {Virga}}, \
  and\ \bibinfo {author} {\bibfnamefont {G.~E.}\ \bibnamefont {Durand}},\
  }\href@noop {} {\bibfield  {journal} {\bibinfo  {journal} {Phys. Rev. E}\
  }\textbf {\bibinfo {volume} {67}},\ \bibinfo {pages} {061701} (\bibinfo
  {year} {2003})}\BibitemShut {NoStop}%
\bibitem [{\citenamefont {de~Gennes}\ and\ \citenamefont
  {Prost}(1993)}]{deGennes}%
  \BibitemOpen
  \bibfield  {author} {\bibinfo {author} {\bibfnamefont {P.~G.}\ \bibnamefont
  {de~Gennes}}\ and\ \bibinfo {author} {\bibfnamefont {J.}~\bibnamefont
  {Prost}},\ }\href@noop {} {\emph {\bibinfo {title} {The Physics of Liquid
  Crystals}}},\ \bibinfo {edition} {2nd}\ ed.\ (\bibinfo  {publisher} {Oxford
  University Press},\ \bibinfo {address} {Oxford},\ \bibinfo {year}
  {1993})\BibitemShut {NoStop}%
\bibitem [{\citenamefont {Straley}(1974)}]{Straley1974}%
  \BibitemOpen
  \bibfield  {author} {\bibinfo {author} {\bibfnamefont {J.~P.}\ \bibnamefont
  {Straley}},\ }\href@noop {} {\bibfield  {journal} {\bibinfo  {journal} {Phys.
  Rev. A}\ }\textbf {\bibinfo {volume} {10}},\ \bibinfo {pages} {1881}
  (\bibinfo {year} {1974})}\BibitemShut {NoStop}%
\bibitem [{\citenamefont {Lipkin}\ and\ \citenamefont
  {Oxtoby}(1983)}]{lipkin83}%
  \BibitemOpen
  \bibfield  {author} {\bibinfo {author} {\bibfnamefont {M.~D.}\ \bibnamefont
  {Lipkin}}\ and\ \bibinfo {author} {\bibfnamefont {D.~W.}\ \bibnamefont
  {Oxtoby}},\ }\href@noop {} {\bibfield  {journal} {\bibinfo  {journal} {J.
  Chem. Phys.}\ }\textbf {\bibinfo {volume} {79}},\ \bibinfo {pages} {1939}
  (\bibinfo {year} {1983})}\BibitemShut {NoStop}%
\bibitem [{\citenamefont {Singh}(2000)}]{singh2000}%
  \BibitemOpen
  \bibfield  {author} {\bibinfo {author} {\bibfnamefont {S.}~\bibnamefont
  {Singh}},\ }\href@noop {} {\bibfield  {journal} {\bibinfo  {journal} {Phys.
  Rep.}\ }\textbf {\bibinfo {volume} {324}},\ \bibinfo {pages} {107} (\bibinfo
  {year} {2000})}\BibitemShut {NoStop}%
\bibitem [{\citenamefont {Miyajima}\ \emph {et~al.}(1990)\citenamefont
  {Miyajima}, \citenamefont {Nakazawa}, \citenamefont {Niikura}, \citenamefont
  {Ujiiye}, \citenamefont {Yashiro},\ and\ \citenamefont
  {Chiba}}]{Miyajima1990}%
  \BibitemOpen
  \bibfield  {author} {\bibinfo {author} {\bibfnamefont {S.}~\bibnamefont
  {Miyajima}}, \bibinfo {author} {\bibfnamefont {K.}~\bibnamefont {Nakazawa}},
  \bibinfo {author} {\bibfnamefont {K.}~\bibnamefont {Niikura}}, \bibinfo
  {author} {\bibfnamefont {Y.}~\bibnamefont {Ujiiye}}, \bibinfo {author}
  {\bibfnamefont {M.}~\bibnamefont {Yashiro}}, \ and\ \bibinfo {author}
  {\bibfnamefont {T.}~\bibnamefont {Chiba}},\ }\href@noop {} {\bibfield
  {journal} {\bibinfo  {journal} {Liq. Cryst}\ }\textbf {\bibinfo {volume}
  {8}},\ \bibinfo {pages} {707} (\bibinfo {year} {1990})}\BibitemShut {NoStop}%
\bibitem [{\citenamefont {Pardhasaradhi}\ \emph {et~al.}(2013)\citenamefont
  {Pardhasaradhi}, \citenamefont {Latha}, \citenamefont {Prasad}, \citenamefont
  {Rani}, \citenamefont {Alapati},\ and\ \citenamefont
  {Pisipati}}]{Pardhasaradhi2013}%
  \BibitemOpen
  \bibfield  {author} {\bibinfo {author} {\bibfnamefont {P.}~\bibnamefont
  {Pardhasaradhi}}, \bibinfo {author} {\bibfnamefont {D.~M.}\ \bibnamefont
  {Latha}}, \bibinfo {author} {\bibfnamefont {P.~V.~D.}\ \bibnamefont
  {Prasad}}, \bibinfo {author} {\bibfnamefont {G.~P.}\ \bibnamefont {Rani}},
  \bibinfo {author} {\bibfnamefont {P.~R.}\ \bibnamefont {Alapati}}, \ and\
  \bibinfo {author} {\bibfnamefont {V.~G. K.~M.}\ \bibnamefont {Pisipati}},\
  }\href@noop {} {\bibfield  {journal} {\bibinfo  {journal} {J. Therm. Anal.
  Calorim.}\ }\textbf {\bibinfo {volume} {11}},\ \bibinfo {pages} {1483–1490}
  (\bibinfo {year} {2013})}\BibitemShut {NoStop}%
\bibitem [{\citenamefont {$\textrm{De Matteis}$}, \citenamefont {Bisi},\ and\
  \citenamefont {Virga}(2007)}]{Virga2007stability}%
  \BibitemOpen
  \bibfield  {author} {\bibinfo {author} {\bibfnamefont {G.}~\bibnamefont
  {$\textrm{De Matteis}$}}, \bibinfo {author} {\bibfnamefont {F.}~\bibnamefont
  {Bisi}}, \ and\ \bibinfo {author} {\bibfnamefont {E.~G.}\ \bibnamefont
  {Virga}},\ }\href@noop {} {\bibfield  {journal} {\bibinfo  {journal} {Con.
  Mech. Therm.}\ }\textbf {\bibinfo {volume} {19}},\ \bibinfo {pages} {1}
  (\bibinfo {year} {2007})}\BibitemShut {NoStop}%
\bibitem [{\citenamefont {Matteis}\ and\ \citenamefont
  {Virga}(2005)}]{Matteis2005b}%
  \BibitemOpen
  \bibfield  {author} {\bibinfo {author} {\bibfnamefont {G.~D.}\ \bibnamefont
  {Matteis}}\ and\ \bibinfo {author} {\bibfnamefont {E.~G.}\ \bibnamefont
  {Virga}},\ }\href@noop {} {\bibfield  {journal} {\bibinfo  {journal} {Phys.
  Rev. E}\ }\textbf {\bibinfo {volume} {71}},\ \bibinfo {pages} {061703}
  (\bibinfo {year} {2005})}\BibitemShut {NoStop}%
\bibitem [{\citenamefont {Turzi}\ and\ \citenamefont
  {Sluckin}(2013)}]{turzi2013}%
  \BibitemOpen
  \bibfield  {author} {\bibinfo {author} {\bibfnamefont {S.~S.}\ \bibnamefont
  {Turzi}}\ and\ \bibinfo {author} {\bibfnamefont {T.~J.}\ \bibnamefont
  {Sluckin}},\ }\href@noop {} {\bibfield  {journal} {\bibinfo  {journal} {SIAM
  J. App. Math}\ }\textbf {\bibinfo {volume} {73}},\ \bibinfo {pages} {1139}
  (\bibinfo {year} {2013})}\BibitemShut {NoStop}%
\bibitem [{\citenamefont {Bates}\ and\ \citenamefont
  {Luckhurst}(1999)}]{bates99}%
  \BibitemOpen
  \bibfield  {author} {\bibinfo {author} {\bibfnamefont {M.}~\bibnamefont
  {Bates}}\ and\ \bibinfo {author} {\bibfnamefont {G.}~\bibnamefont
  {Luckhurst}},\ }\href@noop {} {\bibfield  {journal} {\bibinfo  {journal} {J.
  Chem. Phys.}\ }\textbf {\bibinfo {volume} {110}},\ \bibinfo {pages} {7087}
  (\bibinfo {year} {1999})}\BibitemShut {NoStop}%
\bibitem [{\citenamefont {Pizzirusso}\ \emph {et~al.}(2011)\citenamefont
  {Pizzirusso}, \citenamefont {Savini}, \citenamefont {Muccioli},\ and\
  \citenamefont {Zannoni}}]{pizzirusso11}%
  \BibitemOpen
  \bibfield  {author} {\bibinfo {author} {\bibfnamefont {A.}~\bibnamefont
  {Pizzirusso}}, \bibinfo {author} {\bibfnamefont {M.}~\bibnamefont {Savini}},
  \bibinfo {author} {\bibfnamefont {L.}~\bibnamefont {Muccioli}}, \ and\
  \bibinfo {author} {\bibfnamefont {C.}~\bibnamefont {Zannoni}},\ }\href@noop
  {} {\bibfield  {journal} {\bibinfo  {journal} {J. Mater. Chem.}\ }\textbf
  {\bibinfo {volume} {21}},\ \bibinfo {pages} {125} (\bibinfo {year}
  {2011})}\BibitemShut {NoStop}%
\bibitem [{\citenamefont {Palermo}\ \emph {et~al.}(2013)\citenamefont
  {Palermo}, \citenamefont {Pizzirusso}, \citenamefont {Muccioli},\ and\
  \citenamefont {Zannoni}}]{palermo2013}%
  \BibitemOpen
  \bibfield  {author} {\bibinfo {author} {\bibfnamefont {M.~F.}\ \bibnamefont
  {Palermo}}, \bibinfo {author} {\bibfnamefont {A.}~\bibnamefont {Pizzirusso}},
  \bibinfo {author} {\bibfnamefont {L.}~\bibnamefont {Muccioli}}, \ and\
  \bibinfo {author} {\bibfnamefont {C.}~\bibnamefont {Zannoni}},\ }\href@noop
  {} {\enquote {\bibinfo {title} {An atomistic description of the nematic and
  smectic phases of 4-n-octyl-4' cyanobiphenyl (8cb)},}\ } (\bibinfo {year}
  {2013}),\ \bibinfo {note} {to be published}\BibitemShut {NoStop}%
\bibitem [{\citenamefont {Ferrarini}\ \emph {et~al.}(1996)\citenamefont
  {Ferrarini}, \citenamefont {Luckhurst}, \citenamefont {Nordio},\ and\
  \citenamefont {Roskilly}}]{Ferrarini1996}%
  \BibitemOpen
  \bibfield  {author} {\bibinfo {author} {\bibfnamefont {A.}~\bibnamefont
  {Ferrarini}}, \bibinfo {author} {\bibfnamefont {G.~R.}\ \bibnamefont
  {Luckhurst}}, \bibinfo {author} {\bibfnamefont {P.~L.}\ \bibnamefont
  {Nordio}}, \ and\ \bibinfo {author} {\bibfnamefont {S.~J.}\ \bibnamefont
  {Roskilly}},\ }\href@noop {} {\bibfield  {journal} {\bibinfo  {journal} {Liq.
  Cryst.}\ }\textbf {\bibinfo {volume} {21}},\ \bibinfo {pages} {373} (\bibinfo
  {year} {1996})}\BibitemShut {NoStop}%
\bibitem [{\citenamefont {Luckhurst}(2001)}]{Luckhurst2001}%
  \BibitemOpen
  \bibfield  {author} {\bibinfo {author} {\bibfnamefont {G.~R.}\ \bibnamefont
  {Luckhurst}},\ }in\ \href@noop {} {\emph {\bibinfo {booktitle}
  {{P}hysical~{P}roperties~of~{L}iquid~{C}rystals ~{N}ematics}}},\ \bibinfo
  {editor} {edited by\ \bibinfo {editor} {\bibfnamefont {D.~A.}\ \bibnamefont
  {Dunmur}}, \bibinfo {editor} {\bibfnamefont {A.}~\bibnamefont {Fukuda}}, \
  and\ \bibinfo {editor} {\bibfnamefont {G.~R.}\ \bibnamefont {Luckhurst}}}\
  (\bibinfo  {publisher} {INSPEC},\ \bibinfo {year} {2001})\BibitemShut
  {NoStop}%
\end{thebibliography}%

\end{document}